\documentclass[draft]{agujournal2019}
\usepackage{url} 
\usepackage{lineno}

\usepackage{amsmath}
\newcommand{\xx}{\mathbf{x}}
\newcommand{\uu}{\mathbf{u}}
\newcommand{\vvv}{\mathbf{v}}
\newcommand{\grav}{\mathbf{g}}
\newcommand{\aLAD}{\mathrm{aLAD}}
\newcommand{\aLADtd}{\mathrm{aLAD}_{2D}}
\newcommand{\LADtd}{\mathrm{LAD}_{2D}}
\newcommand{\iDivtd}{\mathrm{iDiv}_{2D}}
\newcommand{\iDiv}{\mathrm{iDiv}}
\newcommand{\dudt}{\frac{\mathrm{D}\uu}{\mathrm{D}t}}

\usepackage{siunitx}
\usepackage{soul}
\usepackage{caption}

\draftfalse

\journalname{}

\begin{document}

%
%


\title{Turbulent Snow Transport and Accumulation: New Reduced-Order Models and Diagnostics}

%
%




\authors{Nikolas O. Aksamit\affil{1}, Alex P. Encinas-Bartos\affil{2}, Holt Hancock\affil{3}, Alexander Prokop\affil{4,5}}


\affiliation{1}{Institute for Mathematics and Statistics, UiT - The Arctic University of Norway, Troms\o, 9037, Norway}
\affiliation{2}{Institute for Mechanical Systems, ETH-Zürich, Zürich, 8092, Switzerland}
\affiliation{3}{Norwegian Geotechnical Institute, Troms\o, 9037, Norway}
\affiliation{4}{University of Vienna, Department for Geology, Vienna, 1010, Austria}
\affiliation{5}{Snow Scan GmbH, Vienna, Austria}


\correspondingauthor{Nikolas O. Aksamit}{nikolas.aksamit@uit.no}

%
%

%
%


\begin{abstract}

Understanding and modeling snow particle dynamics in the atmosphere remains a significant challenge for atmospheric scientists, hydrologists, and glaciologists. Temporally and spatially varying rates of snow transport, deposition, and erosion are driven by atmospheric turbulence and further complicated by inertial particle dynamics. Even with perfectly resolved wind fields, accurately predicting the fate of mobile snow particles in wind relies on semi-empirical assumptions embedded in diffeo-integro equations that contain numerical instabilities. The present research couples a modern approach to snow particle drag with model order reduction tools from nonlinear dynamical systems. Coupled with novel accumulation diagnostics, we provide a simplified framework of snow transport with well-defined simplification errors and rigorous physical meaning. 
\linebreak
\linebreak
Keywords: Snow, Preferential Deposition, Particle Transport, Blowing Snow, Suspension
\end{abstract}

\section{Introduction}

Snow cover dynamics govern the landscape in high latitude and high altitude regions for much of the year. Seasonal snow covers up to 31\% of the Earth's land surface at any time \cite{IPCC2013Chapter4} with up to 40\% of the world's population depending on snow melt for drinking water \cite{Meehl2007}. Snow cover is a consistently evolving medium through melt, snow metamorphism, redistribution, and precipitation. Complex wind-snow coupling that drives the latter two processes results in considerable spatial heterogeneity during winter. Even slight changes in the temporal or spatial variability of snow accumulation can have significant impacts on regional hydrology, ecology, glaciology, and societal safety. For example, spatial heterogeneity modifies basin scale water availability and runoff rates, extent of usable nesting grounds, glacier and ice sheet growth and ablation, and snow avalanche hazards \cite{Schweizer2003, Mott2008,Adam2009, Dadic2010,Callaghan2011, Cimino2025}. In complex terrain, this heterogeneity is amplified by terrain-induced flow structures in near surface winds and orographic effects.

Modeling rates of snow accumulation and erosion requires accurately representing coupled wind-snow processes across a wide range of spatial and temporal scales. At sub-millimeter scales, particle-to-particle bonds govern the erodibility of the snow surface \cite{Schmidt1980} while individual particle geometries, sizes, and densities dictate a non-zero slip velocity of snow traveling in the wind. On the order of meters, wind-generated snow streamers and snow waves influence aperiodic transport rates \cite{Nishimura2024}. At the hundred-meter scale, terrain-generated flow structures, such as ridgetop flow separations and valley channelization influence seasonal transport, accumulation, and sublimation, while kilometer-scale mountain range dynamics and mesoscale meteorology influence the generation of wind and seeder-feeder processes. Due to the multiscale nature of wind-snow coupling, it is currently unfeasible to physically model all the processes governing snow deposition and erosion at all relevant temporal and spatial scales.

As such, in the hydrological and cryospheric sciences, snow accumulation and erosion modeling has been classically approached via semi-empirical relationships. Surface transport rates (i.e. saltation and creep) are typically represented via functions of friction velocity that were adapted from sand studies in flat terrain \cite{Pomeroy1990}. Such approaches struggle to represent unsteady snow transport away from flat terrain or in nonstationary flows \cite{Aksamit2017, Aksamit2018}. Models of snow transport further away from the ground via suspension and preferential deposition typically assume snow particles travel at the same velocity as the wind \cite<see, e.g.,>{Pomeroy1993} or that snow concentrations can be represented as a diffusive medium via empirically-derived advection-diffusion equations, potentially with a vertical settling velocity offset \cite<see, e.g.,>{Lehning2004, Lehning2008, Vionnet2014, Marsh2020}. Such simplifications of inertial particle transport disagree with the physical reality of airborne snow dynamics.

For example, the meteorology and multiphase fluid dynamics communities have made considerable advances in studying and understanding the degree to which velocities of frozen hydrometeors (and generic inertial particles) differ from the surrounding wind. Studies of particle slip velocity that have been performed in laboratories and throughout the atmosphere have generated multiple models of settling variability and descriptions of how snow motions decouple from the wind. Particle and turbulence characteristics have been connected to the non-constant rate at which individual particle settling velocities are enhanced or dampened from expected calm air terminal velocities through physical mechanisms, such as fast-tracking and preferential concentration \cite{Wang1993, Good2014, Nemes2017, Tom2019, Singh2023}. Furthermore, through computationally expensive Lagrangian particle modeling, the recent findings by \citeA{Comola2019} and \citeA{Salesky2019} have highlighted the critical importance of particle inertia on the dominance of windward versus lee-side accumulation patterns, an effect currently neglected in all major snow models.

The present research proposes a new methodology to describe wind-snow coupling, spatially and temporally varying snow transport, and to predict erosion and deposition patterns. Starting with a comprehensive model of forces acting on an inertial particle in a turbulent flow, we utilize tools from nonlinear dynamical systems to simplify snow particle velocity equations to a perturbation of the underlying wind field. This reduced order model (ROM) provides a simplified snow velocity field with well-defined expansion errors. Our ROM faithfully represents multiscale particle dynamics, from the sub-millimeter to kilometer scale. At the individual particle scale, we accurately predict terminal velocities, and trends in loitering and sweeping behavior. We combine our ROM with tools from Lagrangian transport barriers to efficiently predict particle accumulation patterns in two experiments; one spanning seconds and hundreds of meters, and a broader week-long kilometer-scale alpine field experiment. Our ROM is multiple orders of magnitude more efficient than inertial-Lagrangian particle modeling while still capturing its true dynamic complexity.

To accurately and efficiently capture the quickly accelerating environmental change in alpine and polar regions,  we need high-resolution predictions of cryospheric variables \cite{Hock2019HighMountain, Meredith2019PolarRegions}. Wind models continue to improve in accuracy and efficiency, and we must match their fidelity with physics-based models of cryospheric processes \cite{Mott2018}. Our new approach to wind-snow coupling allows us to meet the needs of society at the fidelity necessary to provide subgrid-scale insights into the cryosphere. This success is bolstered by increased computational efficiency that will allow  timely insights at reduced cost.

\section{Methods}

Our approach to predicting spatial and temporal patterns of snow accumulation relies on two new advances for the study of wind-snow coupling. The first component is our new (reduced order) model of snow particle velocities. The Maxey-Riley (MR) equation \cite{Maxey1983} describes the position and velocity of a small, perfectly spherical inertial particle (finite size and mass) in an unsteady turbulent flow as a system of coupled nonlinear differential equations. As detailed below, this equation includes a more comprehensive balance of forces than is common in Lagrangian snow transport studies \cite{Nishimura1998,GrootZwaaftink2014, Okaze2018}, with limited empiricism. This limited empirical control forces the physics of the model, and the physical parameters of the particle and fluid, to define the terminal velocity without user intervention. After adapting the MR equation for fluid and particle densities typical of snow, we derive a reduced order model (ROM) of snow particle velocities using tools from geometric singular perturbation theory \cite{Fenichel1979}. 

Our ROM provides a computationally efficient system of equations and avoids numerically instabilities of the full MR equations. Avoiding this numerical instability is especially beneficial for our second major advance, new Lagrangian and Eulerian snow accumulation diagnostics. These diagnostics are derived from the underlying fluid physics and define the location of maximally attracting or repelling flow structures in snow particle velocity fields. Our quantifiers of snow accumulation and erosion avoid much of the ambiguity of terrain-based accumulation parameterizations \cite{Winstral2002a, Vionnet2021}, while also accounting for more complex inertial dynamics. We detail these two new tools below, with complete mathematical derivations in the Appendix.

\subsection{Large Slip Velocity Maxey-Riley Formulation}
The original MR equation describes the position ($\mathbf{x}$) and velocity ($\mathbf{v}$) of a small spherical inertial particle by balancing multiple forces acting on a particle by the surrounding fluid:

\begin{equation}
\begin{split}
\dot{\mathbf{x}} =&\mathbf{v} \\
\rho_p \dot{\mathbf{v}}= & \rho_f \frac{D\mathbf{u}}{Dt} \\
& + (\rho_p -\rho_f) {\bf g} \\
&-\frac{9 \nu_f \rho_f}{2r^2} (\mathbf{v} - \mathbf{u} - \frac{r^2}{6}\Delta \mathbf{u}) \\
&- \frac{\rho_f}{2}(\dot{\mathbf{v}} - \frac{D}{Dt}(\mathbf{u}+\frac{r^2}{10}\Delta\mathbf{u})) \\
& -\frac{9 \nu_f \rho_f}{2r} \sqrt{\frac{\nu_f}{\pi}}\int_0^t\frac{1}{\sqrt{t-s}}(\dot{\mathbf{v}}(s)-\frac{d}{ds}(\mathbf{u}+\frac{r^2}{6}\Delta\mathbf{u}))ds \\
\label{Eq: Maxey-Riley Original}
\end{split}
\end{equation}
where $\mathbf{u}$ is the temporally and spatially varying wind velocity, $\frac{D}{Dt}$ is the material derivative, $\grav$ denotes the gravitational acceleration, $\rho_f$ and $\rho_p$, are representative densities of the air and particle, respectively, $r$ is the radius of the particle, $\nu_f$ is the kinematic viscosity of the air, and $t$ and $s$ are time variables. The terms on the right-hand side include the buoyancy effect, the Stokes drag on the particle, the force from the fluid moving with the particle, and the Basset-Boussinesq memory term, respectively. The terms containing $r^2\Delta\uu$ are commonly referred to as the Fauxén corrections \cite{Haller2008}. Equation (\ref{Eq: Maxey-Riley Original}), however, assumes that linear (Stokes) drag approximates the drag force and that the particle Reynolds number 

\begin{equation}
    Re_p=\frac{|\vvv-\uu|2r}{\nu_f},
    \label{Eq: Re_p}
\end{equation}
is much less than one. The mass and relatively high density of snow particles when compared with air results in a significant slip velocity that negates the Stokes regime assumption, though this linear drag model is still sometimes used in Lagrangian snow modeling \cite{Hames2022, Hames2025}. To account for more complex drag situations, small particle to wind length scale ratios, and larger particle slip velocities, we introduce the modified MR equation:

\begin{align}
\dot{\xx}&=\vvv, \nonumber\\
V\rho_p\dot{\vvv}&=V \left[\rho_f \dudt + \left(\rho_p-\rho_f \right)\grav - \frac{\rho_f}{2}\left(\dot{\vvv} -\dudt \right) \right] -\frac{A}{2}\rho_f C_D|\vvv-\uu|(\vvv-\uu).
\label{Eq: MR General}
\end{align}
where $V$ is the particle volume, $A$ is the cross-sectional area of the particle experiencing drag, and $C_D$ is a non-constant drag function that depends on the particle wake and slip velocity (see the Appendix for more details). To obtain equation (\ref{Eq: MR General}) from equation (\ref{Eq: Maxey-Riley Original}), we assume that the particle is very small with respect to the length scales of the fluid, in which case the Fauxén corrections are negligible. The coefficient of the Basset-Boussinesq memory term can be expressed as the product of the Stokes drag coefficient, $\frac{9\nu_f\rho_f}{2r^2}$ and $\frac{r}{\sqrt{\pi\nu_f}}$. Following common practices in the existing literature \cite{Michaelides1997,Haller2008}, we further assume that $\frac{r}{\sqrt{\nu_f}}$ is also very small, such that we can neglect the integral from the original Maxey-Riley equation. With equation (\ref{Eq: MR General}), we can utilize either spherical or non-spherical particle geometries. For the time being, we will proceed with a spherical particle assumption, but similar analysis on the non-spherical geometry transport equations can be found in the Appendix.

To account for a wide range of $Re_p$ beyond the Stokes regime, we use the $Re_p$-dependent drag function $C_D$ from \citeA{Abraham1970}:

\begin{equation}
C_D=C_0\left(1+ \frac{\delta_0}{\sqrt{Re_p}}\right)^2.
\label{Eq: C_D}
\end{equation}
where $C_0$ and $\delta_0$ are experimentally derived constants. This drag formulation is commonly applied for frozen hydrometeors studies \cite<see, e.g.,>{Bohm1989, Heymsfield2010}, with empirically-derived coefficients defined for snow and ice particles. Equation (\ref{Eq: C_D}) has great flexibility with the small $Re_p$ limit recreating the Stokes drag while the large $Re_p$ limit generates a constant drag, $C_0$.

For our initial investigations, we model snow particles of volume $V$ as spheres with an effective diameter, $d_{eff}$ such that $V=\frac{\pi}{6}d_{eff}^3$. This also results in a calculation of effective particle density much smaller than the density of ice and is more representative of what has been recently measured in nature \cite{Rees2021, Singh2021}. We can further simplify our equations of pseudo-sphere motion as 
\begin{equation}
\dot{\vvv}=\beta \dudt + (1-\beta)\grav - \frac{\beta}{2d} C_D|\vvv-\uu|(\vvv-\uu),
\label{eq: Dimensional Snow}
\end{equation}
where $\beta = \frac{3\rho_f}{2\rho_p+\rho_f}$.

With this particle velocity model (\ref{eq: Dimensional Snow}) in hand, we seek to harness recent developments in Lagrangian coherent structures and transport barriers to understand temporally and spatially varying wind features that attract or repel snow particles in transport. Lagrangian coherent structures act as the backbone of the flow \cite{Haller}, with their inertial counterparts governing snow concentrations and organizing wind-snow coupling. Unfortunately, numerically integrating (\ref{eq: Dimensional Snow}) directly has a significant computational burden, as is discussed more in the Results. Furthermore, Lagrangian diagnostics of accumulation are most effectively generated by integrating particle trajectories backwards in time from a grid of final locations where accumulation information is desired. Due to the separated timescales (distinct slow and fast systems) governing inertial particle motion \cite<see, e.g.,>{Haller2008}, there is a finite-time numerical instability when performing backwards integration of particles. That is, small numerical errors grow exponentially fast in backward time. To avoid this computational burden and numerical instability, we adapt the slow manifold simplifications of \citeA{Haller2008} for our nonlinear drag case, and reduce the snow particle velocity equations (\ref{eq: Dimensional Snow}), in non-dimensional form, to a perturbation of the underlying non-dimensional wind field:

\begin{equation}
\begin{split}
\vvv&=\uu + \alpha^4 \uu_4+ \alpha^7 \uu_7 \\ 
\mathbf{u}_4&=\left[\frac{2Re}{\beta C_0\delta_0^2} \right](1-\beta)(\grav-\dudt) \\
\uu_7&=-\frac{2\sqrt{Re}}{\delta_0}\sqrt{|\uu_4|}\uu_4.
\label{eq: MR ROM}
\end{split}
\end{equation}
Here all terms have been scaled by the characteristic length scale $L$, and velocity scale $U$, and $\alpha=\sqrt{d/L}$. See the Appendix for full details of the derivation of our reduced order model (\ref{eq: MR ROM}). This ROM approximates the velocity of a snow particle as a perturbation to the underlying fluid velocity, and one can in turn predict asymptotic snow transport and accumulation at an accuracy analogous to the underlying wind field, within suitable ranges of $\alpha$. This avoids the arbitrary choice of initial particle velocity in Lagrangian snow modeling, as occurs when solving equation (\ref{Eq: MR General}) directly. Furthermore, analysis of the velocity field (\ref{eq: MR ROM}) provides a means to directly utilize well-established Lagrangian and Eulerian coherent structure theory to understand transport dynamics, as described in the next section.

\subsection{Accumulation Diagnostics}
\label{Sec: Accumulation Diagnostics}
With our snow particle velocity field (\ref{eq: MR ROM}), we are able to apply tools from nonlinear dynamics to identify accumulation zones. Consider a landscape defined laterally by $(x,y)$ coordinates, with the land's height above sea level defined as a function $z=f(x,y)$. We seek to quantify the amount of snow that accumulates on the surface during a time window $t\in[t_0, t_1]$ after falling through the three-dimensional time-varying wind field $\uu(x,y,z,t)$. For the moment, we will disregard redistribution of snow particles once they come in contact with the ground (i.e., sticky particles). 

 For a given particle trajectory $\xx(t)$, parameterized by time $t$ with initial position $\xx_0=\xx(t_0)$, and velocity $\vvv(\xx(t),t)$, the particle number concentration in a small volume around $\xx(t)$, call it $n(\xx(t))$, evolves over time as 
 
 \begin{equation}
     n(\xx(t_1))=n(\xx(t_0))e^{-\int_{t_0}^{t_1}\nabla\cdot\vvv(\xx(t),t)dt}.
     \label{eq: Number Density Evo}
 \end{equation}
 For  $t_0<t_1$, \citeA{Oettinger2018}, via an equivalent formulation for a different inertial model, showed that regions of large {\it negative} Lagrangian averaged divergence (LAD),
 
 \begin{equation}
     \mathrm{LAD}^{t_0,t_1}(\xx_0)=\frac{1}{|t_1-t_0|}\int_{t_0}^{t_1}\nabla\cdot\vvv(\xx(t),t)dt,
 \end{equation}
where $\nabla\cdot\vvv=\partial\vvv_1/\partial\xx_1 + \partial\vvv_2/\partial\xx_2 +\partial\vvv_3/\partial\xx_3$, identifies $t_0$ positions of finite time attractors of inertial particles and quantifies the average rate of airborne particle accumulation along particle paths. In direct analogy, by reversing the path of integration and traveling backwards from $\xx(t_1)$ to $\xx(t_0)$, we can identify the $t_1$ position of finite time attractors as regions of large {\it positive} LAD.

For $t\in[t_0, t_1]$, particles starting at various heights may actually come to rest at the same position on the boundary of our flow for different $t_0<t<t_1$. Thus, for a given position on the boundary of a flow domain, $\xx_1=\xx(t_1)$ we can quantify the cumulative impact of in-air particle convergence to particles that land at $\xx_1$ by calculating the LAD for all trajectories that come to rest at $\xx_1$ between $t_0$ and $t_1$. We achieve this with the {\it accumulated} LAD,

\begin{equation}
\aLAD^{t_1,t_0}(\xx_1)= \frac{1}{|t_0-t_1|}\int_{t_1}^{t_0}\mathrm{LAD}^{t_1,s}(\xx_1)ds=\frac{1}{|t_0-t_1|}\int_{t_1}^{t_0}\left[\frac{1}{|s-t_1|}\int_{t_1}^s\nabla\cdot\vvv(\xx(t),t)dt \right]ds.
\label{eq: aLAD}
\end{equation}

\begin{figure}
    \centering
    \includegraphics[width=0.5\linewidth]{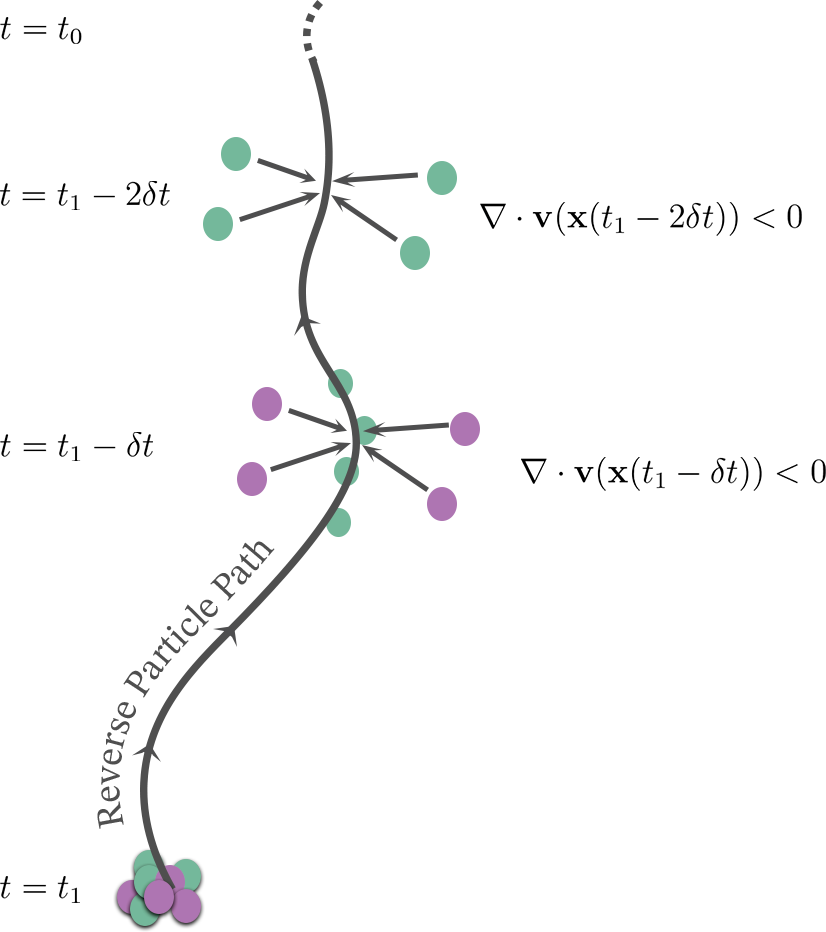}
    \caption{Relating backward integration of a snow particle trajectory (black) to accumulation of nearby particles (green and purple). Negative divergence corresponds with nearby particle moving toward from our trajectory (horizontal arrows). With sufficient settling time, accumulation of particles to the trajectory in forward time results in a high concentration at $t=t_1>t_0$ on the ground, and large negative $\aLAD$.}
    \label{fig:aLAD}
\end{figure}

 Large negative $\aLAD^{t_0,t_1}(\xx_1)$ values indicate significant attraction of nearby particles to a path that comes to rest at the boundary location $\xx_1$. This process and calculation is visualized in Figure \ref{fig:aLAD}. From a final resting position at time $t=t_1$, the divergence is numerically calculated along the particle path at multiple timesteps ($t_1-\delta t$, $t_1-2\delta t$,$\dots$), and summed over all potential snowfall times between $t_0$ and $t_1$ to account for ground particles that originated at different heights.
 
Our initial investigations of inertial particle dynamics in atmospheric LES reveal that falling particles often collect along vertically oriented layers of high shear, in line with previous studies on preferential concentrations \cite{Wang1993}. This results in a large vertical component of Lagrangian divergence oriented along the fall direction. This vertical divergence initially appears to be less relevant than lateral divergence for ground accumulation patterns over time, especially with heavier particles. This may be because lateral convergence to coherent structures occurs quickly, and vertically aligned particles subsequently accumulate on top of each other. The underlying physical process is a topic of future research, but this behavior appears consistent throughout the following experiments. As a first order measure of spatial ground accumulation patterns, we focus on the lateral divergence of the inertial particle field. Lateral divergence can be thought of as an inverted inertial variation of surface clustering found with floating particles \cite<e.g.,>{Lovecchio2013}. The rate of lateral accumulation (i.e., collection of particles in the gravity-normal plane) can be objectively quantified with the accumulated 2D Lagrangian averaged divergence:
\begin{equation}
\aLADtd^{t_1,t_0}(\xx_1)=\frac{1}{|t_0-t_1|}\int_{t_1}^{t_0}\left[\frac{1}{|s-t_1|}\int_{t_1}^s\nabla_{2D}\cdot\vvv(\xx(t),t)dt \right]ds.
\label{eq: aLADproj}
\end{equation}
where $\nabla_{2D}\cdot\vvv=\partial\vvv_1/\partial\xx_1 + \partial\vvv_2/\partial\xx_2$. We can analogously define $\LADtd$ for lateral in-air accumulation. Figure \ref{fig:aLAD} demonstrates how we quantify lateral accumulation to an attracting trajectory, and potential vertical divergence along attracting trajectories (e.g. the vertical spread of green particles at time $t=t_1-\delta t$).

The formulation of equations (\ref{eq: aLAD}) and (\ref{eq: aLADproj}) also provide a well-defined instantaneous limit as $t_0\to t_1$. In fact, the instantaneous rates of accumulation of boundary (ground) particles in two and three dimensions are simply

\begin{equation}
    \begin{split}
        \lim_{t_0\to t_1}\aLAD^{t_1,t_0}(\xx_1)&=i\mathrm{Div}^{t_1}(\xx_1)=\nabla\cdot\vvv(\xx_1,t_1) \\
        \lim_{t_0\to t_1}\aLADtd^{t_1,t_0}(\xx_1)&=i\mathrm{Div}^{t_1}_{2D}(\xx_1)=\nabla_{2D}\cdot\vvv(\xx_1,t_1).
        \label{eq: Instantaneous aLAD}
    \end{split}
\end{equation}
The instantaneous limits (\ref{eq: Instantaneous aLAD}) are the divergence of the {\it inertial} particle velocity field (i.e., $\iDiv$) and provide an Eulerian snapshot of particle accumulation at a given position and time. When calculated over a surface of positions immediately above the ground, equation (\ref{eq: Instantaneous aLAD}) quantifies the influence of near surface wind structures on particle accumulation and dispersion using a perturbation of the underlying wind field that only requires physical properties of the particle and fluid.

Recent years have seen a rapid advancement in cold regions hydrological modeling platforms for complex terrain, but hydrologists and cryospheric scientists continue to struggle with accurate predictions of snow accumulation zones \cite{Mott2018}. The most common approach to account for accumulation in sheltered areas is the terrain-based $S_x$ parameter from \citeA{Winstral2002a}. For a given point on the terrain, $S_x$ calculates the maximum rate of elevation change from that point to all points in a search domain defined by the predominant wind direction. After selecting user-defined parameters, $S_x$ provides a binary categorization of ground-sheltering behind terrain features. The subsequent rate of accumulation or erosion in sheltered or unsheltered areas is then prescribed by the atmospheric or blowing snow models chosen by the user \cite<e.g.,>{Schon2015,Schon2018}. As a common basis of understanding, we also compare the spatial patterning predicted by our flow-based diagnostics against $S_x$ for an Arctic-alpine study site over a one-week period of high precipitation. Additionally, we conducted terrestrial laser scanning (TLS) measurements of the spatial snow depth distribution at our study site to validate our accumulation diagnostics results in natural winter conditions \cite<e.g.,>{Mott2010,Schneiderbauer2011,Vionnet2014,Prokop2016}. We followed the methodology described by \citeA{Prokop2008,Prokop2009} and \citeA{Prokop2009b}. 

\subsection{Particle Statistics}
\label{Section: Particle Stats}
For our numerical accumulation tests, we modeled ground-truth particle trajectories using the modified MR equation (\ref{eq: Dimensional Snow}). Particle densities and diameters were varied, depending on the experiment (defined in the corresponding section). In Section \ref{Section: Sweep Loiter}, we compare probability density functions of MR particle dynamics with ROM (eq. \ref{eq: MR ROM}) dynamics at equivalent particle densities and diameters. In Section \ref{Sec: Cube Flow}, we compare particle concentrations from multiple brute-force numerical advection simulations against the accumulation diagnostics defined in Section \ref{Sec: Accumulation Diagnostics}. All accumulation diagnostics (e.g., $\iDiv$, $\aLAD$, etc.) are calculated using the inertial particle velocity field (\ref{eq: MR ROM}). In order to compare concentrations of sparse, chaotically advected particles with gridded diagnostics, we binned particle positions at specified times in two-dimensional horizontal grids. These grids were consistently of dimension $200\times80$ to mimic the resolution of the underlying velocity field. Specific characteristics of particles in each bin where then analyzed, as described in the specific tests below. All correlations reported in Section \ref{Sec: Cube Flow} are statistically significant with a p-value less than $10^{-10}$.

\subsection{Wind Flow Data}
\label{Sec: Wind Data}
\begin{figure}
    \centering
    \includegraphics[width=\linewidth]{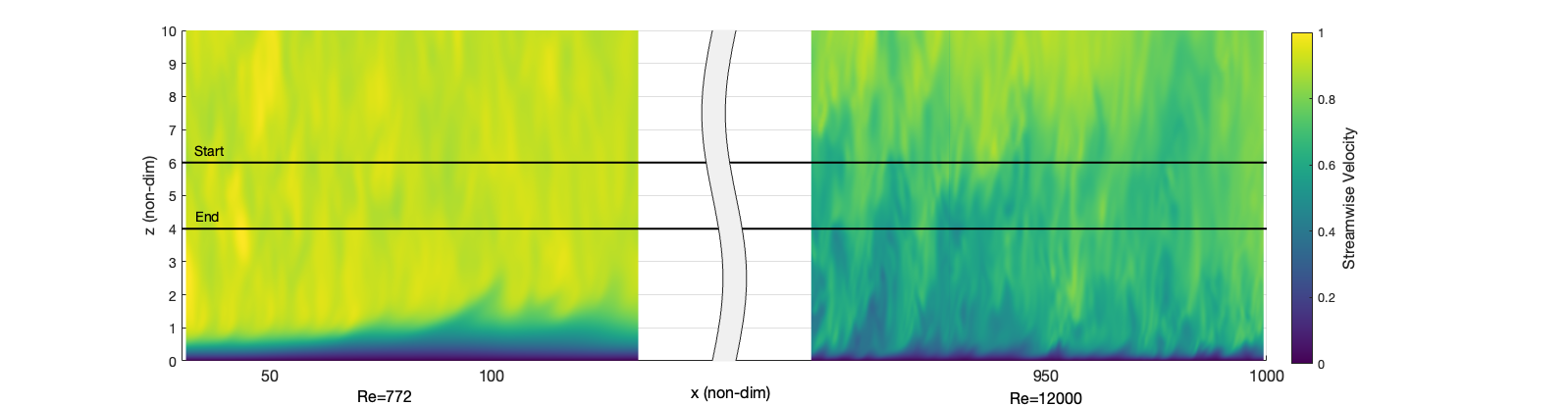}
    \caption{Example streamwise-vertical slices of instantaneous streamwise velocity for two subsets of the Johns Hopkins Turbulence Database Transitional boundary layer dataset revealing low and high Reynolds number flow structure. Start and approximate finish height of falling inertial particles studied in Section \ref{Section: Sweep Loiter} also indicated.}
    \label{fig: JHTDB DNS}
\end{figure}

We use two sources of fluid flow data to evaluate inertial particle dynamics at distinct spatial scales. To test the individual particle scale dynamics in turbulent and non-turbulent conditions, we rely on direct numerical simulations (DNS) of a transitional boundary layer from the Johns Hopkins Turbulence Database \cite{Perlman2007,Li2008,Zaki2013}. This DNS calculates the development of a turbulent boundary layer in an incompressible fluid over a flat plat with an elliptical leading edge (Figure \ref{fig: JHTDB DNS}). The available data covers a spatial domain of $x\in[30.2185,1000.065]\hat{L}$, $y\in[0,240]\hat{L}$, and $z\in[0.0036,26.4880]\hat{L}$, and temporal domain $t\in[0,1175]L/U_{\infty}$, with $\hat{L}$ being a dimensionless DNS length unit, and $U_\infty$ the free-stream velocity. At the beginning of the domain, the boundary layer thickness Reynolds number is approximately 772, and increases to over 12,000 at downstream locations. We advect particles in two subdomains, $x\in[31,131]\hat{L}$ and $x\in[900,1000]\hat{L}$, to evaluate if our MR and ROM models represent the sweeping and loitering behavior described in other turbulent inertial-particle studies \cite<e.g.,>{Nielsen1993, Good2014}. 

Additionally, to understand the meter and kilometer scale particle concentration patterns predicted by our ROM and accumulation diagnostics, we developed two different neutral atmosphere turbulent large eddy simulations using the PALM platform \cite{Maronga2020}. Our first experiment involves a turbulent flow around a cube with a mean flow velocity of 2 ms$^{-1}$. The cube is 40 meters on all sides, centered laterally in a 400 m $\times$ 180 m domain, with leading edge 260 m downstream from the inflow. We ran a six-hour flat terrain precursor run to generate turbulence for PALM's turbulent inflow generator \cite{Maronga2020}, and then ran an additional six hours of cube flow simulation, studying only the last 1000 s. We then analyze the accumulation patterns in the turbulent wake behind the cube, using trajectories from the Maxey-Riley equation (\ref{eq: Dimensional Snow}) to evaluate small particle proxies from the ROM (\ref{eq: MR ROM}) with LAD, $\aLAD$, and $\iDiv$. 

Our second PALM simulation was designed for an Arctic-alpine study site northwest of Troms\o, Norway, using 5 nested domains. For the ground digital elevation model, we used the 1 meter resolution national height model of Norway \cite{Kartverket2014}. The largest flow domain utilized a 120 m $\times$ 120 m horizontal grid spacing, with 20 m vertical spacing, spanning 73$\times$57$\times$5 km. This allowed sufficient development of large scale forcing and more realistic upstream flow dynamics in the highly mountainous region. The grid nesting also included 40 m, 20 m, 10 m, and 5 m domains with two-way coupling. The innermost 5 m domain spanned the TLS scan area at 2.4$\times$2.4$\times$1.1 km. As with the cube flow, a six-hour precursor run was performed to generate turbulence to be utilized in PALM's turbulent inflow generator. A pure southwesterly geostrophic surface velocity was set at 2.8 $ms^{-1}$ for five different six-hour runs to generate a small ensemble of flows mimicking the atmospheric conditions suggested by meteorological data during the period. Details of both PALM simulations can be found in Table \ref{tab:simulation_params}.

\begin{table}[h]
    \centering
    \begin{tabular}{|l|c|c|}
        \hline
        Parameter & Cube Flow & Ullstinden \\ 
        \hline
        Domain Size & 400$\times$180$\times$80 m & Outermost: 73$\times$57$\times$5 km, Innermost: 2.4$\times$2.4$\times$1.1 km \\ 
        Spatial Resolution & 2$\times$2$\times$2 m & Outermost: 120$\times$120$\times$20 m, Innermost: 5$\times$5$\times$5 m \\ 
        Run Time & Precursor + 2 h & Precursor + 6 h ($\times$ 5)\\ 
        Investigation Period & 1000 s & 500 s \\ 
        Turbulence Generator & Noncyclic/Turbulence Inflow & Noncyclic/Turbulence Inflow \\ 
        \hline
    \end{tabular}
    \caption{PALM LES Simulation parameters for Cube Flow and Ullstinden cases.}
    \label{tab:simulation_params}
\end{table}

\section{Results}
It is worth remembering that an exhaustive experimental validation of equations (\ref{eq: Dimensional Snow}) or (\ref{eq: MR ROM}) is not currently feasible. To test these models at the individual particle level requires high spatial and temporal resolution measurements of both the particle velocity, as well as the velocity and derivatives of the surrounding fluid. Such data has not been collected for snow, and is sparsely available for any medium \cite{Brandt2022}. Instead, we test the performance our particle equations against multiple particle behavior benchmarks from the individual particle scale (<mm) up to the mountain scale (km).

\subsection{Individual Particle Terminal Velocities}
\label{Sec: Terminal Velocity}
To start, we calculate the terminal velocities predicted for a range of spherical particle sizes and densities, and compare with an empirically-validated function of particle Reynolds number that is common in snow-turbulence studies \cite{Bohm1989,Heymsfield2010,Nemes2017,Singh2021,Singh2023,Li2024}. Based on arguments of dimensional analysis \cite{Davies1945}, we can estimate the terminal velocity $(v_t)$ of a perfect spherical particle in a still fluid via the Best-Davies number ($X$) as

\begin{equation}
v_t = \frac{\eta Re_p}{d\rho_f}, \qquad Re_p=\frac{\delta_0^2}{4}\left[\left(1+\frac{4\sqrt{X}}{\delta_0^2\sqrt{C_0}}\right)^{\frac{1}{2}} -1 \right]^2, \qquad X=\frac{\rho_f}{\eta^2}\frac{8m\grav}{\pi} \label{eq: Best-Davies}
\end{equation}
where $\eta$ is the dynamic viscosity of the air. Following the findings of \citeA{Heymsfield2010}, we use $C_0=0.6$ and $\delta_0=5.83$ for the remainder of our study. These values also appear in our ROM (\ref{eq: MR ROM}). We can test the terminal velocities of particles in still air as predicted by the modified MR equation by setting $\uu=0$ in (\ref{eq: Dimensional Snow}) and solving for the fixed point when acceleration ceases, $\dot{\vvv}=0$.

\begin{figure}
    \centering
    \includegraphics[width=0.75\linewidth]{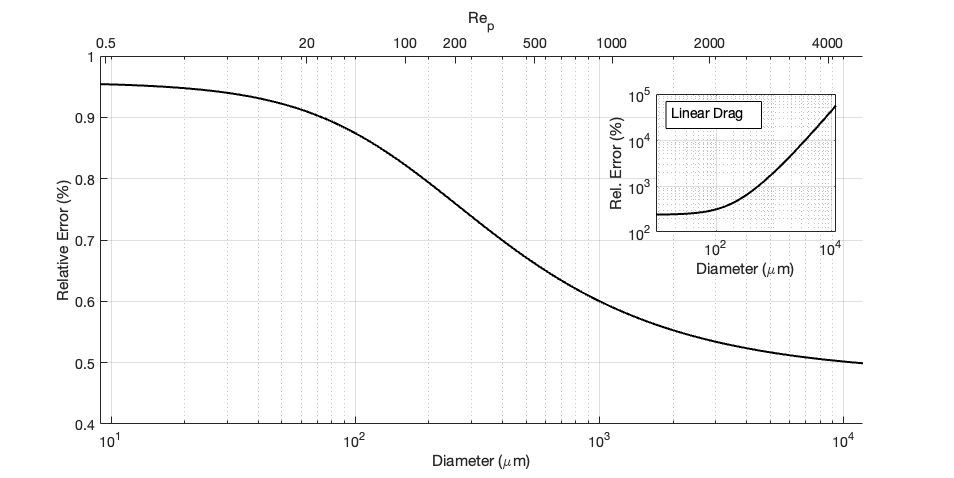}
    \caption{Relative error of terminal velocities of modified MR from Best-Davies terminal velocity calculations for a range of particle Reynolds numbers that encompasses and extends well beyond snow particle ranges. Inset shows relative error of original MR equation with linear drag for same range of values.}
    \label{fig: MR to Best}
\end{figure}

In Figure \ref{fig: MR to Best}, we show the relative error in terminal velocities calculated with our modified MR equation (\ref{eq: Dimensional Snow}) against the Best-Davies approach (\ref{eq: Best-Davies}) for a wide range of particle diameters at a fixed particle density corresponding to spherical graupel (140 kg m$^{-3}$) \cite{Ishizaka1993,Li2023}. The particle diameter and resultant $Re_p$ are both plotted against the relative error. Figure \ref{fig: MR to Best} indicates that our nonlinear drag modification allows us to estimate the terminal velocities of spherical particles with less than 1\% errors up to $Re_p=4000$, well beyond the Stokes drag regime required in the original MR equation. In fact, the accuracy of our approximation increases as particle $Re_p$ increases.  In contrast, the inset in Figure \ref{fig: MR to Best} shows the same error calculations using the linear drag MR formulation, which obtains errors of $10^4$\% at these high $Re_p$ values. This indicates that our modified MR equation is accurately predicting the terminal velocity, a crucial aspect of hydrometeor behavior and, for spherical particles, supports the dimensional analysis arguments of \citeA{Davies1945}. We thus use the nonlinear MR equation as a ground truth for particle motion in several validations of the more computationally efficient ROM.

\begin{figure}
    \centering
    \includegraphics[width=\linewidth]{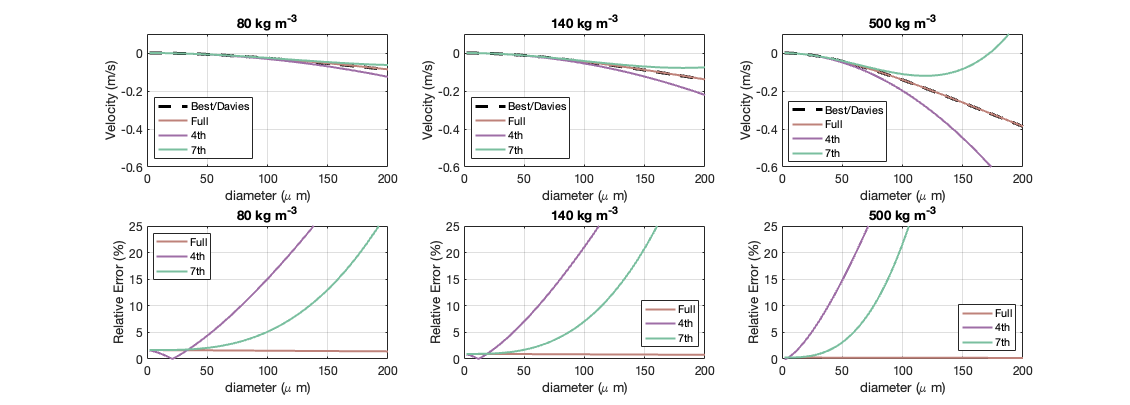}
    \caption{Top Row: Terminal Velocities of MR, reduced order approximations, and "Best" (Davies) terminal velocity calculations for a range of particle densities and diameters. Bottom Row: Relative error of terminal Velocities of modified MR, reduced order approximations, and "Best" (Davies) terminal velocity calculations for the same range of particle densities and diameters.}
    \label{fig: ROM Terminals}
\end{figure}

By setting $\uu=0$ in the ROM (\ref{eq: MR ROM}), we can perform a similar terminal velocity test. In Figure \ref{fig: ROM Terminals}, we let our individual snow particle densities range from 80 kg m$^{-3}$ to 500 kg m$^{-3}$. This corresponds to the range of observed values for freshly fallen aggregates, dendrites, needles, and graupel from \citeA{Ishizaka1993, Ishizaka2016, Rees2021, Singh2021, Li2023} and \citeA{Singh2024}. We compare the terminal velocities and relative errors of a lowest order approximation (including $\uu_4$) and a higher order approximation (with $\uu_4$ and $\uu_7$) in equation (\ref{eq: MR ROM}) against the empirical Best-Davies approach. For reference, we also include a comparison between the Best-Davies values and MR (\ref{eq: Dimensional Snow}).

As our ROM is a perturbation solution of the MR equations, we see an expected decrease in ROM accuracy as our diameters and perturbation parameter $\alpha=\sqrt{d/L}$ increases.  The seventh order approximation in equation (\ref{eq: MR ROM}) approximates the terminal velocity with less than a 25\% error up to nearly 200 $\mu$m, 150 $\mu$m, and 100 $\mu$m for the 80, 140 and 500 kg m$^{-3}$ cases, respectively. A 25\% relative error is on the same order as the relative error of previous comparisons with experimental data \cite{Bohm1989, Heymsfield2010}.

While falling snow experiments, especially imagery-based studies, often measure snow particles with effective diameters above this range of high ROM accuracy ($>200\mu m$), the small particles we effectively represent are common in both surface snow transport and clouds. For example, in a review of 10 different blowing snow studies, \citeA{Gordon2009} found average particle diameters for snow in transport were less than 200 $\mu m$. As the measured diameters fit a two-parameter gamma distribution, diameters less than the mean accounted for up to 60\% of the particles. Furthermore, diameters of ice crystals in clouds extend well below 20 $\mu m$ \cite{Platt1997}. As such, the physics described by these ROMs has immediate and accurate applications for simplifying many fields of frozen hydrometeor research. 

When predicting accumulation in nature, however, we cannot control the size distribution of the snow particles we seek to describe. In Section \ref{Section: Deposition Patterns}, we ask the question, how well do predictions of small snow particle transport represent transport patterns of relatively larger particles? We subsequently show that inertial dynamics defined by small particle proxies provide an accurate starting point for representing wind-snow coupling and accumulation at both meter and kilometer scales, in both numerical and field studies.

\subsection{Collective Sweeping and Loitering of Particles}
\label{Section: Sweep Loiter}
Slightly increasing the spatial scale of interest, we evaluate the modeled collective behavior of groups of particles in turbulent flows. Many benchmark theoretical, numerical, and experimental studies have sought to predict transitions in the qualitative influence of turbulent flows on falling particle velocities (see \citeA{Brandt2022} for a recent review). One such influence is the apparent acceleration or deceleration of falling particles from their predicted still air terminal velocities due to the presence of turbulence. Sweeping is the process by which heavy particles are preferentially accelerated, whereas loitering describes particles that fall slower than expected. These processes have been attributed to clustering and accumulation, fast-tracking, and various other mechanisms, depending on the size of flow structures and individual particle characteristics. We rely on a direct numerical simulation of a turbulent boundary layer where the finest scales of fluid motion are resolved to test our particle transport model. In this way, individual particle speeds are not influenced by sub-grid parameterizations.

The Johns Hopkins Turbulence Database transitional boundary layer \cite{Perlman2007,Li2008,Zaki2013} provides a perfect test bed to validate sweeping and loitering aspects of our models at varying levels of turbulence. Due to the fine resolution of the DNS method, it is currently unfeasible to run a DNS at Reynolds numbers comparable to an atmospheric boundary layer ($Re\approx10^8$). This means the ratio of inertial and viscous forces is much smaller than for a natural snow study. As such, we account for the different flow parameters by adjusting our particle parameters. To mimic a consistent viscous resistance at the particle scale while respecting the lower Re values in the DNS, we non-dimensionalize our particle parameters differently than in the atmosphere. That is, for a particle diameter of $50\mu m$, we scale by $L\in\{0.1m,1.48m\}$ and $U=0.12ms^{-1}$ to mimic an air viscosity of $\nu=1.48\times10^{-5}m^2s^{-1}$ while matching the Re at the front and the back of the flow, respectively. This indicates the largest scale structures in this DNS would be much smaller than what one would find in the atmosphere, as expected, but provides realistic fluid resistance to particles at relevant densities and diameters.

In both regions, we initiate 160,000 falling particles at $z=6\hat{L}$ on a 100 $\hat{L}$ $\times$ 50 $\hat{L}$ grid and let them fall for 20 model timesteps. In the $Re=772$ domain, this height (see Figure \ref{fig: JHTDB DNS}) lies just above the top of the boundary layer ($z\approx \hat{L}$), whereas in the $Re=12000$ domain, particles from $z=6\hat{L}$ are transported within the turbulent boundary layer which extends to $z\approx 15\hat{L}$. This allows us to evaluate the different behavior of particles interacting with larger scale structures, and different degrees of turbulence intensity.

\begin{figure}
    \centering
    \includegraphics[width=\linewidth]{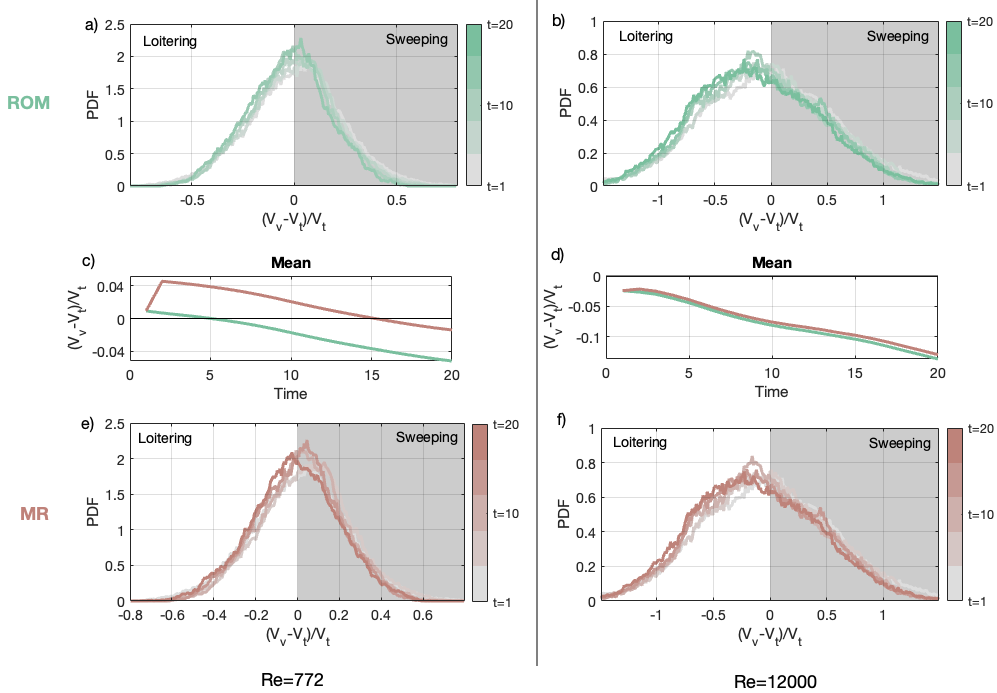}
    \caption{Temporal evolution of relative velocity PDFs for 160,000 inertial particles falling in a transitional boundary layer DNS. Particles trajectories and velocities were calculated with the reduced order model and modified Maxey-Riley equation in both low and high Reynolds number sections. Distribution averages are plotted over time in panels c and d. All flows had a tendency towards loitering, with increased loitering as the particles were organized by flow structures.}
    \label{fig: DNS Sweeping-Loitering}
\end{figure}

Figure \ref{fig: DNS Sweeping-Loitering} summarizes the temporal evolution of the relative velocity, $(V_v-V_t)/V_t$ for four experiments with simulated falling particles. We calculate $V_v$ as the instantaneous vertical velocity of each particle, and $V_t$ is the still-air terminal velocity from the MR equation. These probability distributions in Figure \ref{fig: DNS Sweeping-Loitering} are akin to the recent snow study of \citeA{Singh2023} (their Figure 4), but for particles of a single diameter. Panels a and b show the temporal evolution of the PDF of relative velocity as simulated by the reduced order model over twenty DNS time steps in the Re=772 and Re=12000 domains, respectively. Panels e and f show analogous temporal PDF evolutions for the full MR model, while c and d show the mean relative velocity over all particles over time for the two particle models. The MR experiments used ROM particle velocities as initial conditions.

Due to small changes in particle velocities, MR and ROM particles did not follow the same paths. In all experiments, however, distributions of relative particle velocity exhibited similar trends. In the $Re=772$ region, MR particles exhibited relatively more sweeping, with a mean  positive bias of 4$\%$ in $(V_v-V_t)/V_t$ (Figure \ref{fig: DNS Sweeping-Loitering}c), but a parallel trend towards loitering as time progressed. In the $Re=12000$ region, this mean bias shrinks to less than 0.1$\%$ (Figure \ref{fig: DNS Sweeping-Loitering}d), with a strong dominance of loitering for all particles, and an increase in loitering strength over time. The close match of ROM to MR results is related to the smaller $\alpha$ values at high Re (due to larger $L$), and better constrained error estimates. This is doubly encouraging considering $L$ is orders of magnitude larger in the atmosphere.

The inertial particles evaluated here do not perfectly represent snow due to the parameters chosen to match the DNS, and thus do not reflect snow tendencies towards sweeping \cite<e.g.,>{Singh2023}. Instead, this particle-flow study is more representative of the early loitering work from \citeA{Nielsen1993}. In fact, the loitering exhibited here confirms Nielsen's relative turbulence intensity predictions that loitering can dominate when $\sigma/v_t<3$, where $\sigma$ is turbulence intensity. We find $\sigma/v_t$ values of 1.5 and 1.9, for the low and high Re regions, respectively.

Initial ROM relative velocity distributions are nearly centered around zero for both Re regions, but values span twice the range in the more turbulent flow. This is due to both larger vertical velocity values and derivatives with increased turbulence. As the particles fall, they become more organized by the flow, and tend to loiter more. This is analogous to Good et al.'s (2014) conclusions in isotropic turbulence, where the presence of large scale structures were necessary for loitering. Particles in the $Re=772$ region start above the strongest shear structures, and descend towards developing surface rolls. Starting at the same height in the $Re=12000$ region, particles initially start in a much more turbulent region and their initial degree of loitering is only enhanced as their positions are further organized in the flow. This trend is further supported in the Appendix where results from a comparable advection experiment initiated at $z=3\hat{L}$ are shown. When initiating particles closer to the surface, in both $Re=772$ and $Re=12000$ flows, their interaction with stronger large scale structures results in even more significant loitering.

These findings indicate that the ROM can effectively recreate sweeping and loitering tendencies present in the full MR equation for a generic inertial particle. On a 10-core desktop computer, the reduced order model took approximately 20 seconds to advect 20,000 particles in this turbulent domain, whereas the full Maxey-Riley equation test took 5,000 seconds. This 250x speed-up is exceptional considering the agreement in particle velocity distributions for complex particle behavior in both domains, and nearly trivial differences in the high turbulence cases.

\subsection{Test of Deposition Patterns}
\label{Section: Deposition Patterns}
In the following sections we perform a numerical and field-based validation of the ground accumulation diagnostics $\aLAD^{t_0,t_1}(\xx_1)$ and $\aLADtd^{t_0,t_1}(\xx_1)$, and rates of inertial accumulation $\mathrm{iDiv}^{t_1}(\xx_1)$ and $\iDivtd^{t_1}(\xx_1)$ as calculated with our reduced order particle velocity field. These diagnostics are shown to reveal the natural spatial heterogeneity of snow depth and provide predictive capacity for short-term (seconds) and long-term (days) accumulation and ablation zones. We primarily focus on snowfall dynamics where the terminal velocity of a falling particle is less than the wind speed. Given that the upper end of snowfall speeds are on the order of 1 ms$^{-1}$, and atmospheric and mountain winds often exceed this speed, even close the ground, this is the context most relevant for describing snowfall behavior. Our first accumulation study considers a simple turbulent flow around a bluff body.

\subsubsection{Meter Scale Structure-Influenced Particle Concentrations}
\label{Sec: Cube Flow}
Using the PALM LES platform \cite{Maronga2020}, we generated approximately 15 minutes of spatially resolved neutral atmosphere turbulent flow around a 40 meter cube, temporally sampled at one second intervals (see Section \ref{Sec: Wind Data}). Focusing on a 150 $\times$ 160 $\times$ 80 m region in the turbulent wake, we calculated ten 20-second snowfall experiments of 2.5 million particles each. We simulated snowfall in three experiments using the full MR equation (\ref{eq: Dimensional Snow}) for particles with a fixed density of 140 kg m$^{-3}$, but three different diameters. All particle accumulation diagnostics (LAD, $\aLAD$ and $\iDiv$) were calculated using an ROM velocity field with 140 kg m$^{-3}$ particles and 100 $\mu$ m diameters. MR snowfall particles were initialized on a 3D grid spanning 2 to 4 meters above the ground, and were subsequently organized by the turbulent wake and boundary layer structures. Once the particles hit the ground, their position was frozen. This generated short-term concentration changes as particles traveled in the flow. 

\begin{figure}
    \centering
    \includegraphics[width=0.85\linewidth]{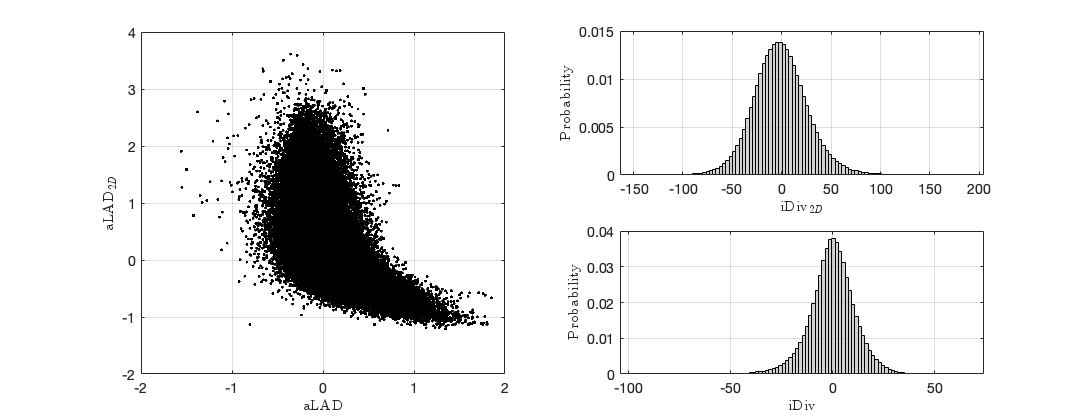}
    \caption{Evidence of lateral convergence of particles and vertical separation of falling particles in Lagrangian and Eulerian lateral and vertical divergence fields for the reduced order model.}
    \label{fig: aLAD_vs_aLAD2D}
\end{figure}

As described in Section \ref{Sec: Accumulation Diagnostics}, falling snow in turbulent flows is often laterally organized by coherent flow structures (e.g., moved to eddy edges), and subsequently concentrated vertically along thin shear interfaces. This is akin to the preferential concentrations observed in inertial particles in general \cite{Wang1993, Baker2017, Voth2017, Tom2019}, and the snow transport heterogeneity evident in snow waves and snow snakes \cite<see, e.g.,>{Nishimura2024}. In Figure \ref{fig: aLAD_vs_aLAD2D} we show the inverse statistical relationship that this concentration behavior has on ROM particles. Panel \ref{fig: aLAD_vs_aLAD2D}a plots the instantaneous $z=0.5$ m $(\aLAD,\aLADtd)$ values for 100 twenty-second advections. We find that strong forward-time stretching of particle clusters along shear lines or surfaces, as indicated by positive $\aLAD$, corresponds with clear lateral convergence (negative $\aLADtd$). In contrast, forward-time lateral divergence of falling particles (positive $\aLADtd$) corresponds with a vertical compression of trajectories  (slightly negative $\aLAD$). Thus, predicting an area of strong lateral concentration near the ground is closely related to finding a region of vertical separation of particles, if given enough time for the vertical column of snow to accumulate.

The underlying fluid flow is incompressible (zero-divergence), but the reduced-order model of particle velocity (\ref{eq: MR ROM}) is a compressible vector field. Figures \ref{fig: aLAD_vs_aLAD2D}b and \ref{fig: aLAD_vs_aLAD2D}c leverage this non-zero divergence to explain the organization of falling snow patterns. The lateral instantaneous divergence ($\iDivtd$) over the entire flow domain is slightly skewed negatively, suggesting particles are preferentially accumulating in specific areas as they fall, thus creating a non-uniform distribution on the ground. The three dimensional divergence $\iDiv$, however, is actually dominated by the vertical divergence in these accumulation zones, skewing the distribution slightly positive. The impact of the vertical particle divergence on ground accumulation depends on the relative lateral advection versus settling velocity. Indeed, laterally attracting features may provide fast-tracking to the ground for heavy particles, while slowing ground accumulation for light particles. 

\begin{figure}
    \centering
    \includegraphics[width=\linewidth]{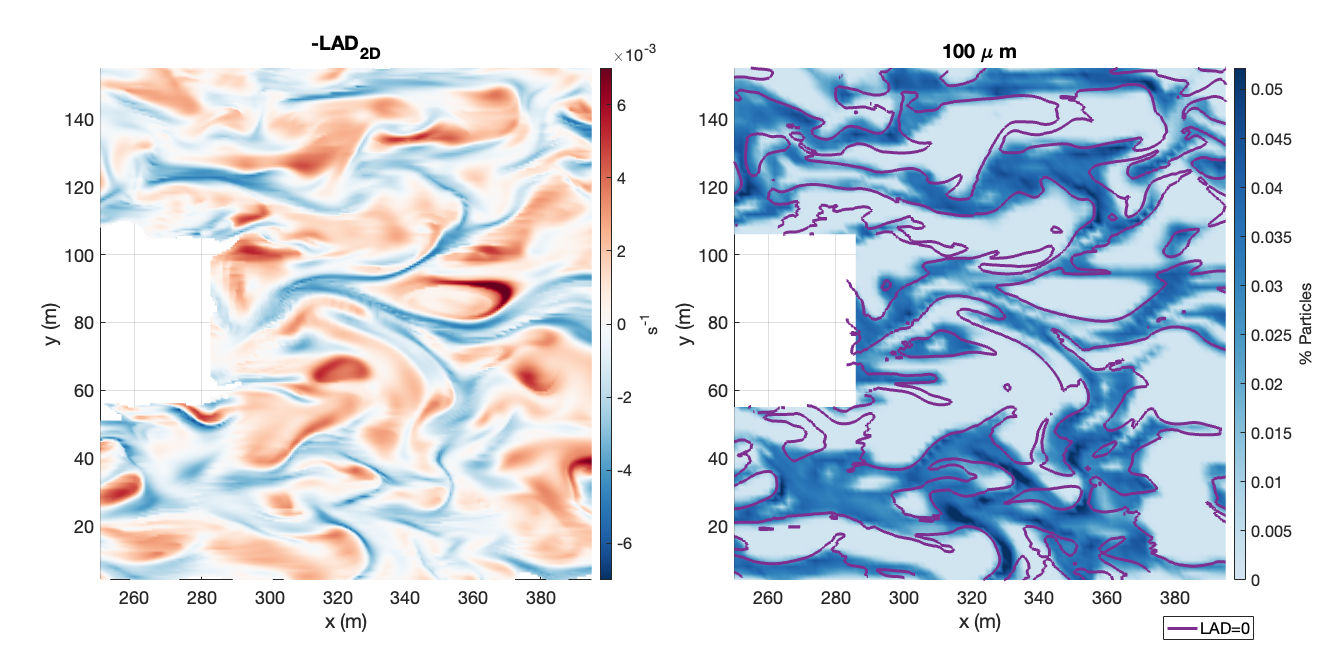}
    \caption{Comparison of $-\LADtd$ at $z=3$ m (left) and percentage of horizontally binned snow particles in the surrounding height band $z\in[2.5,3.5]$ m (right). The $\LADtd=0$ contour is outlined in purple on the right, showing the boundary separating horizontal convergence versus divergence. Note general agreement between high divergence and (red zones, left) with particle free regions (white, right)}
    \label{fig: LAD to MR Mid Air}
\end{figure}

Figure \ref{fig: LAD to MR Mid Air} shows an example of this lateral convergence along flow structures. The left panel displays the $-\LADtd$ field calculated at $z=3$ m for a twenty-second backward time window. Recall, negative values of $-\LADtd$ correspond with strong lateral convergence in the flow, concentrating particles, but also disperse particles vertically, as seen in Figure \ref{fig: aLAD_vs_aLAD2D}. The right panel shows gridded airborne particle concentrations, following Section \ref{Section: Particle Stats}, for particles between $z=2.5$ and $z=3.5$ m. Large white areas indicate regions with no particles, and correspond with large particle divergence predictions from the ROM and LAD (red in Figure \ref{fig: aLAD_vs_aLAD2D}). In fact, the boundaries separating particle divergence and convergence (LAD=0) are drawn in purple, and closely approximate the boundaries of the particle-free regions. Our ROM and the LAD accurately predict the mid-air collection of particles. As we will see, this does not immediately translate to higher ground accumulation on short timescales for all particles.

\begin{figure}
    \centering
    \includegraphics[width=\linewidth]{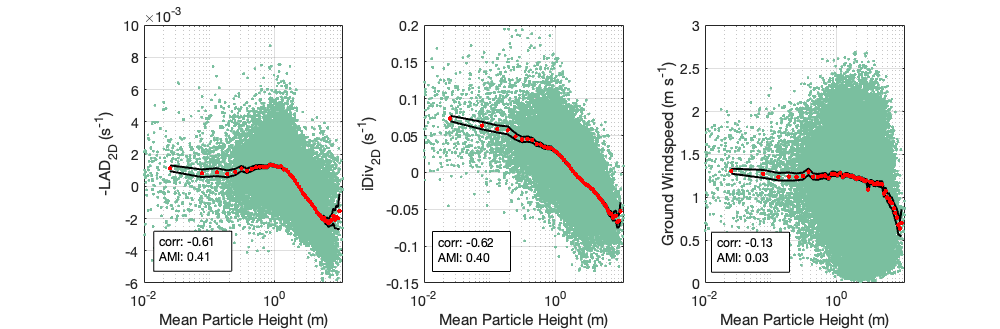}
    \caption{Comparison of $-\LADtd$, $\iDivtd$, and wind speed at $z=3$ m with average height of horizontally binned snow particles. The plot displays total results from ten 20-second experiments with 2.5 million particles each. Large lateral airborne convergence corresponds with short-term vertical separation, as can be seen in strong negative trends between particle height and divergence diagnostics. In contrast, a weak correlation suggests slow ground wind speed is only loosely related to greater particle height. Red dots indicate mean values and 95\% confidence intervals are shown in black.}
    \label{fig: Stats Mid Air}
\end{figure}

Figure \ref{fig: Stats Mid Air} reveals that the regions of high particle concentration in Figure \ref{fig: LAD to MR Mid Air} also correspond with significant vertical travel. We compared bin-averaged particle heights for ten 20-second experiments with the corresponding $-\LADtd$, $\iDivtd$, and wind speed values at $z=3$ m. There is clear evidence that a decrease in $-\LADtd$ and $\iDivtd$ corresponds with greater particles height. That is, the strong lateral convergence to thin shear zones redistributes the particles vertically. This is further supported by a strong negative correlation coefficient (-0.61 and -0.62). There is a less clear trend with wind speed, with a large range of particle heights for each wind speed, and a much weaker correlation. The question remains, how well do these particle convergence diagnostics translate into predicting patterns in ground accumulation?

\begin{figure}
    \centering
    \includegraphics[width=\linewidth]{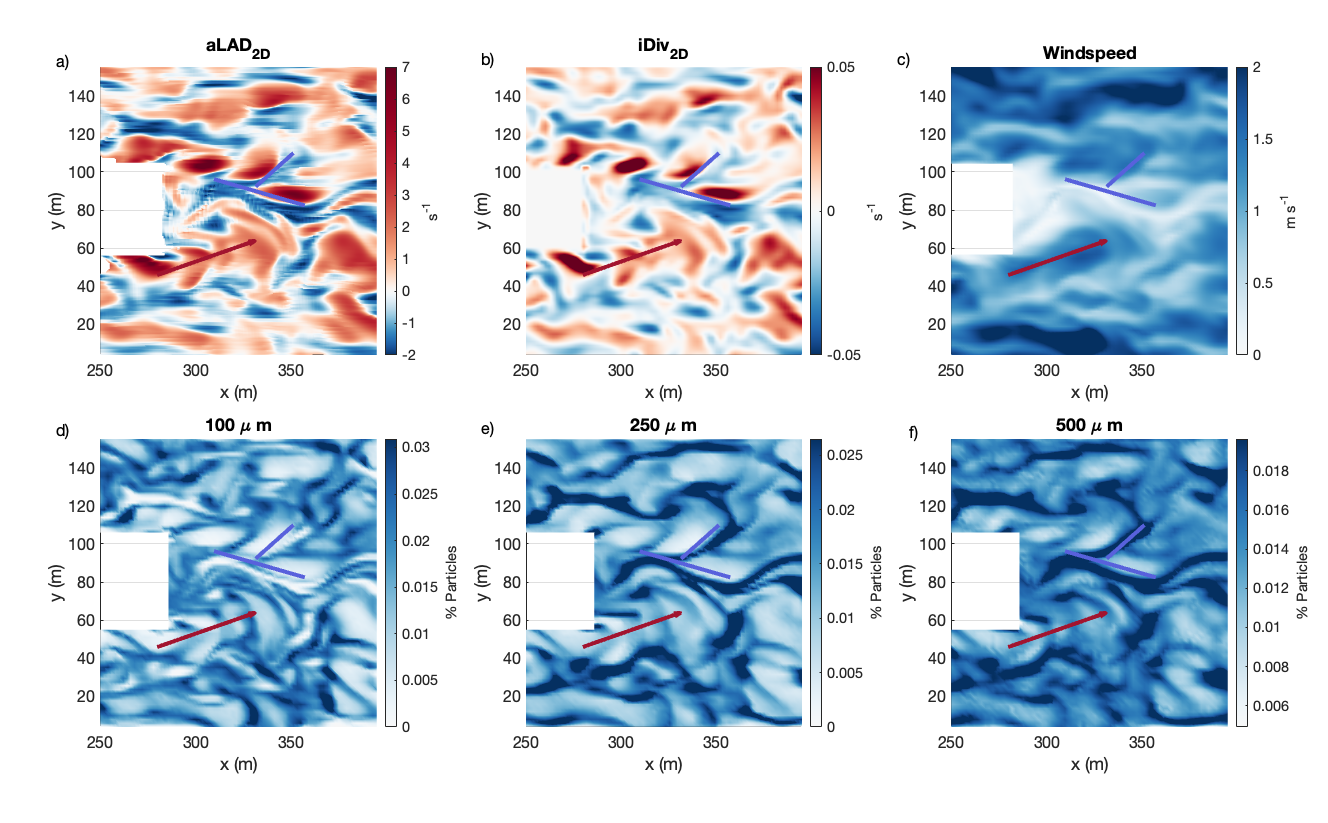}
    \caption{Panels a-c: Three meter wind diagnostics $\aLADtd$, $\iDivtd$, and wind speed at the end of a 20-second snowfall period in a turbulent wake. Panels d-f: ground accumulation of MR-driven snow particles, of various sizes, after 20 seconds of advection. Low concentration plume is highlighted with a red arrow, and accumulation streaks wake are highlighted by blue lines.}
    \label{fig: ROM to MR Vary Sizes}
\end{figure}

Using these same ten 20-second runs, we sought to understand how well accumulation concentrations could be predicted by our ROM and diagnostics, for a variety of particle sizes. To achieve this, we use the same 100 $\mu m$ MR simulations, but also run 250 $\mu m$, and 500 $\mu m$ simulations. These two larger particle sizes corresponded with a poor terminal velocity approximation by our ROM in Section \ref{Sec: Terminal Velocity}. We then binned all particles below 3 m after twenty seconds using the same 2D grid as before. We compare these near-ground accumulations with the ground-based $\aLADtd$, $\iDivtd$, and wind speed, all calculated at $z=0.5$ m. Snapshots from one such experiment are displayed in Figure \ref{fig: ROM to MR Vary Sizes}. We could additionally compare with the terrain-based $Sx$ accumulation parameter \cite{Winstral2002a}, but the simple geometry of the flow and the influence of user-defined parameters to generate $Sx$ reveals little insight; $Sx$ shows a uniform accumulation zone behind the cube, whose dimensions (width and extent) are wholly defined by user parameters.


A slow meandering wake can be seen in the wind speed field (\ref{fig: ROM to MR Vary Sizes}c). We highlight the core of this low wind speed feature with a diagonal blue line behind the cube in all panels ($y\in[80,100]$). A distinct lateral branch is also marked with a blue line in all panels. Large negative values of $\aLADtd$ and $\iDivtd$ overlap with this region as well. For small particles (100  $\mu m$), this quasi-linear feature primarily corresponds with lower particle accumulation numbers, with thin sections of enhanced accumulation. This is contrary to physical intuition that suggests snow should accumulate in low wind speed regions. Similar features can be found around the cube, where other quasi-linear footprints of strong convergence corresponds with thin accumulation bands. We can also effectively predict snow-free patterns, as seen when comparing snow concentration with the large red divergence zones in the flow, such as the tip vortex highlighted with the red arrow in Figure \ref{fig: ROM to MR Vary Sizes}.

The primary remaining complication with predicting accumulation zones using $\aLADtd$ and $\iDivtd$ appears to be the balance of settling velocity and rates of lateral convergence along thin shear zones. When we instead compare the strong convergence zones behind and to the side of the cube for 250 and 500 $\mu m$ particles (panels e and f), we find a clear signature of maximized accumulation. The clear high particle concentration zones match perfectly with the $\aLADtd$ accumulation regions, and closely mimic $\iDivtd$. The low concentration zones for larger particles also match high divergence regions. 

\begin{figure}
    \centering
    \includegraphics[width=\linewidth]{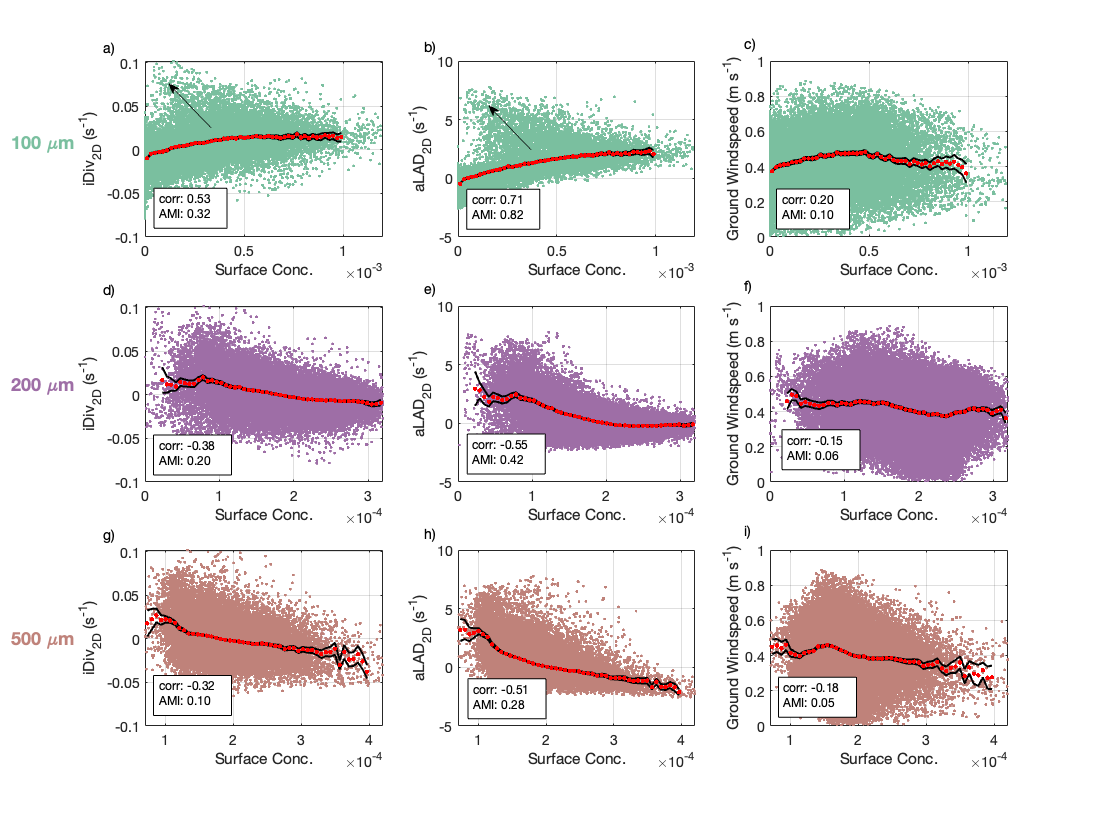}
    \caption{Bulk statistics of near-surface wind diagnostics (L to R: $\iDivtd$, $\aLADtd$, wind speed) against ground particle concentration after 20 seconds of advection in turbulent cube wake for various particle diameters. A general increase in concentration with decreasing $\iDivtd$ and $\aLADtd$ is statistically significant for 200 and 500 $\mu m$ particles. For small particles, vertical dispersion along convergence zones creates the opposite effect on short-term particle accumulation. Weaker, but still statistically significant relationships, between particle concentrations and wind speed exist but provide no predictive use. Red dots indicate mean values and 95\% confidence intervals are shown in black.}
    \label{fig: Cube Stats 100 mu}
\end{figure}

We test the statistical significance of these relationships by calculating the correlation and average mutual information between surface concentration values and adjacent ground diagnostics for our ensemble of simulations. Figure \ref{fig: Cube Stats 100 mu} shows the comparison of $\aLADtd$, $\iDivtd$, and wind speed at 0.5 meters with the relative fraction of particles below 3 meters for all 20 second experiments, and three particle sizes. We calculate $\aLADtd$ over the corresponding 20 second snowfall window. $\iDivtd$, and wind speed are measured at the final 20 second timestep.

For MR simulations of 100 $\mu m$ particles, the short-time suspension of small particles by vertical particle divergence, and lateral organization, is evident in a nonlinear decrease of concentration with decreasing $\iDivtd$ and $\aLADtd$. This is further supported by trends in mean values (red dots) as well as a positive correlation with particle height for both diagnostics (0.53 and 0.71, respectively). There is evidence that some reduced accumulation can be attributed to high divergence, as indicated by the black arrows in panels a and b. As we will show, this effect becomes much stronger with larger particles. In contrast, there is less of a trend present in bulk or mean values of wind speed, resulting in a weaker, but still statistically significant correlation (0.2).

The larger settling velocity of 200 $\mu m$ and 500 $\mu m$ particles dominates any counter-effect of vertical dispersion. The correlations between $\iDivtd$ and $\aLADtd$ ground concentrations switch for these larger particles and a nonlinear decrease is present. The strongest correlations between surface concentrations and either accumulation diagnostic is found with $\aLADtd$. This suggests identifying the flow features affecting small particle organization provides a good approximation of organization of much larger particles, with stronger inertial effects, as well. In contrast, a similar, small magnitude ($\le0.2$) correlation exists between wind speed and ground concentration exists for all  particles.

Over short times, neither high nor low wind speeds can guarantee a certain ground accumulation outcome. At these same timescales, vertical divergence of particles along flow structures can help slow down small particle ground accumulation, and may also provide fast-tracking for heavier particles. As such, lateral divergence diagnostics have proven useful for predicting ground accumulation patterns and rates, with potential improvement after even longer settling times.

\subsubsection{Mountain Scale Accumulation: Ullstinden Study Site}
\label{Section: Ullstinden}
Our final example tests accumulation pattern prediction across larger temporal and spatial scales. Two consecutive terrestrial LiDAR scans were performed surrounding a storm cycle from November 27 to December 4, 2024 at Ullstinden, a 1000 meter peak 30 kilometers northwest of Troms\o, Norway. The scans were performed from an adjacent carpark at 100 meters ASL, and processed to provide gridded snow depth change ($\Delta$Hs) at a square spatial resolution of 3 m, with measurements covering an area of approximately 2.5 km$^2$, and an estimated relative accuracy between 10 and 20 cm. Data below 300 m were filtered due to a forest influence which was not accounted for in the PALM run.

During this observation period, direct field observations from the authors indicated relatively calm winds from the southwest, and a large accumulation of snow. Northern Norway has an extremely sparse network of meteorological stations, especially at high elevation. As such, standardized meteorological observations over the study domain are not currently available. A "Snower" style IoT snow depth station in Oldervik, a sea level village immediately adjacent to Ullstinden, reported a snow depth change of approximately 40 cm between November 27 and December 4 scans, with minimal scouring. Gridded 1 km model wind data from the Norwegian Water Resources and Energy Directorate's snow model seNorge \cite{Saloranta2012, Lussana2019} indicates a general southwesterly flow during the storm, as is typical of large winter precipitation events for the region. This is further supported by 10 meter wind data from ERA5 reanalysis \cite{hersbach2023era5} which suggest average speeds of approximately 2 m/s from south to southwest over the domain. There was a distinct shift in wind direction on December 3 during a lull in the storm during which wind speeds dropped and there was no precipitation. The snowfall resumed on the evening of December 3rd with winds remaining below 1.5 m/s from the S and SW again. These conditions were conducive to preferential deposition and minimal redistribution, as was also observed by several of the authors that were on site.

As described in Section \ref{Sec: Wind Data}, we first generate precursor turbulent conditions, and then simulate five independent six hour PALM runs. Our domain of interest was the innermost of five progressively finer-resolution nested domains, all of which were driven by a static forcing of 2.8 $m/s$ wind from the southwest. Nesting our small high-resolution study site (2.4$\times$2.4 km extent) inside lower resolution surroundings (73$\times$57 km extent) provided suitable upwind coherent flow structure development. 

Calculating Lagrangian snow particle trajectories with meter resolution data over kilometer scale domains is expensive, even with the advantages of a ROM. In light of the success of $\iDivtd$ to predict accumulation and particle height in section \ref{Sec: Cube Flow}, we rely on $\iDivtd$ to test accumulation in our inertial particle models at this Arctic-alpine site.  Velocity fields from the last 100 seconds of each 6 hour run were recorded as output with a one second sampling rate. Near-surface (10 meter) $\iDivtd$ fields were computed for all 500 velocity fields and averaged over time. This approach allowed the development of persistent terrain-induced turbulent structures to form, while averaging dampened the impact of transient near-surface structures. The resulting $\overline{\iDiv}_{2D}$ quantified the long-term impact of terrain-based features that remained quasi-stationary throughout all runs (e.g., ridgeline flow separation, acceleration, channelization), as well as recurrent non-stationary flow features that reappeared in certain regions (e.g., bursting mechanisms).

\begin{figure}
    \centering
    \includegraphics[width=\linewidth]{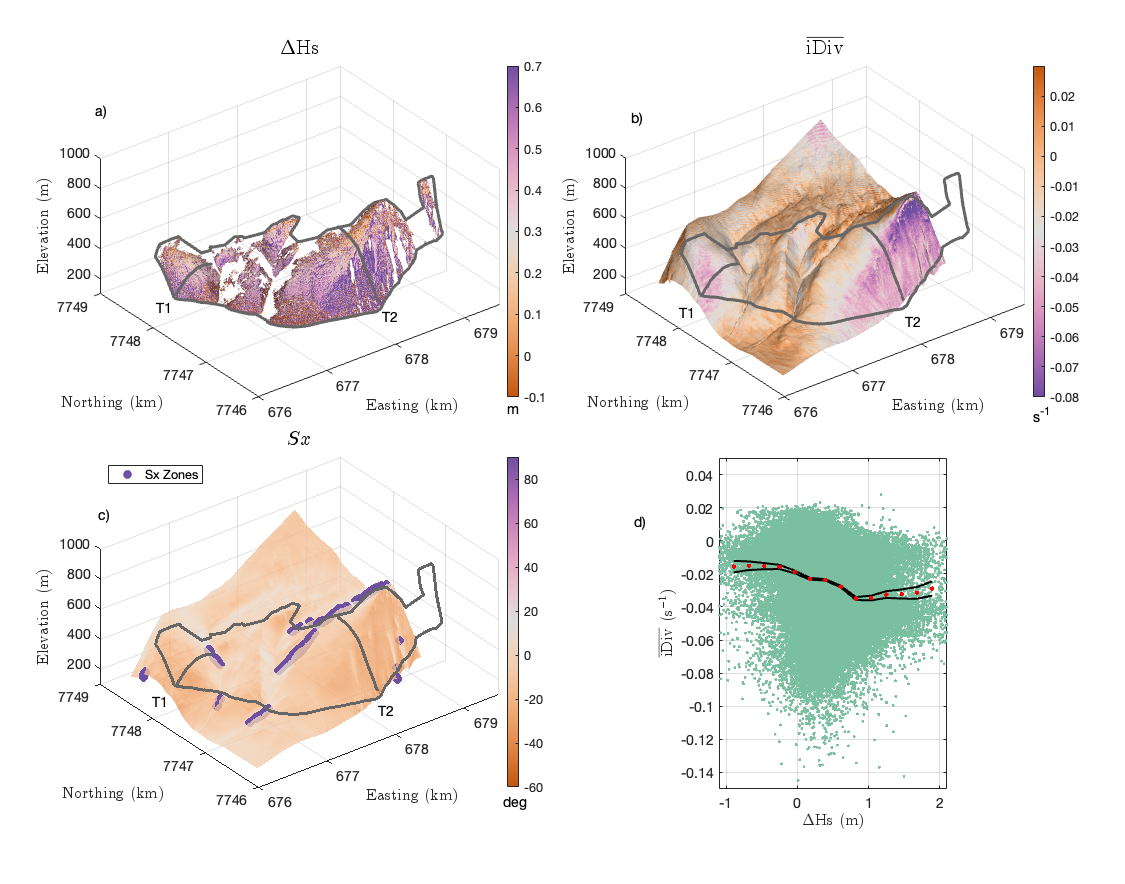}
    \caption{Panel a: One week snow depth change (27.11.24 to 4.12.24) derived from terrestrial LiDAR scanning. Panel b: Average near surface two dimensional divergence of particle velocity field. Panel c: Winstral Sx sheltering parameter with accumlation zones highlighted in purple. Panel d: Bulk and mean relationship between $\iDiv$ and gridded snow depth change over all LiDAR measurements.}
    \label{fig: Ullstinden Compare}
\end{figure}

In Figure \ref{fig: Ullstinden Compare}a we show the change in snow height ($\Delta Hs) $ during the early December storm. Figures \ref{fig: Ullstinden Compare}b-c display $\overline{\iDiv}_{2D}$ and the terrain-based parameter $Sx$, respectively. The boundary of the scan domain is drawn in grey for reference in all diagnostics, and two 50-meter-wide transects are also marked as T1 and T2. Following the previous work on snow accumulation in complex-terrain, we use an $Sx$ search window of 300 meters \cite<e.g.,>{Winstral2017, Vionnet2021} and a sheltering-induced accumulation for values of $Sx>20^\circ$ \cite{Wood1995}. The predominant wind speed is set as southwest for $Sx$ as well, and the upstream search window is 60$^{\circ}$ wide \cite{Winstral2002b}. In Figure \ref{fig: Ullstinden Compare}c, we highlight the above-threshold sheltering zones with purple markings, and color the remaining zones with their $Sx$ values. 

Due to the predominant southwesterly nature of the geostrophic wind flow, $Sx$ predicts sheltering and snow accumulation on the north and east aspects of rises in the terrain. The south and west aspects are expected to experience maximum wind exposure and scouring. The measured snow accumulation patterns do not support this in reality. Dark purple regions on south facing and east facing aspects in panel a indicate large increases in snow depth over the course of the storm, with limited depth change near ridgeline, and some ridgetop scouring. The difference between $Sx$ and actual accumulation is due to several factors, one of which is the upstream redirection of the wind by other mountains in the area. As shown in the cube wake in Section \ref{Sec: Cube Flow}, accumulation cannot be guaranteed by wind speed alone. As well, inertial effects can impact whether particle accumulate on lee or windward aspects \cite{Comola2019}, which is also not accounted for by $Sx$.

Comparing $\Delta Hs$ with $\overline{\iDiv}_{2D}$ reveals a striking similarity. The nested PALM wind-simulations appear to faithfully represent wind conditions at the study site as low $\overline{\iDiv}_{2D}$ regions identified two prominent accumulation zones bisected by transects T1 and T2. These south and west facing aspects experienced cross loading and windward accumulation, which $\overline{\iDiv}_{2D}$ was able to predict. In addition, high $\overline{\iDiv}_{2D}$ predicted divergence of near-surface particles and a greater likelihood of erosion. Figure \ref{fig: Ullstinden Compare}d shows a bulk scatter plot of all snow depth changes and corresponding $\overline{\iDiv}_{2D}$. Mean binned values and their 95$\%$ confidence intervals are shown in red and black respectively. There is a clear trend in increasing $\Delta Hs$ with decreasing $\overline{\iDiv}_{2D}$, though the nonlinear relationship creates significant scatter.

\begin{figure}
    \centering
    \includegraphics[width=0.75\linewidth]{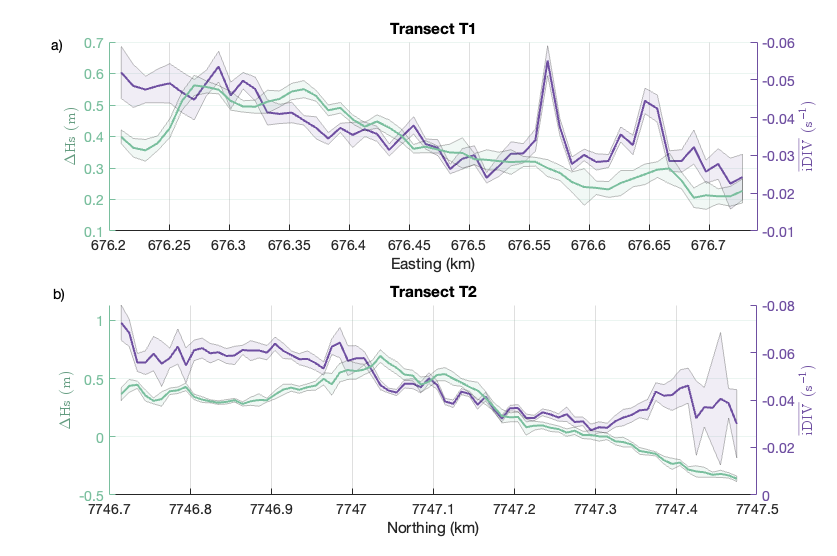}
    \caption{Comparison of $\iDiv$ and one week snow depth change along 50 meter wide transects (T1 and T2) from Figure (\ref{fig: Ullstinden Compare}). Note a general match in trends along transects, though exact ridgeline scouring onset varies.}
    \label{fig: Transect Compare}
\end{figure}

We refine this slope scale comparison to two transects spanning leeward-windward transitions. Figure \ref{fig: Transect Compare}a shows values of $\Delta Hs$ and $\overline{\iDiv}_{2D}$ averaged over a 50 meter wide region following the track T1 in Figure \ref{fig: Ullstinden Compare}. Mean values are shown in bold colors with 95$\%$ confidence intervals lightly shaded. Along T1, there is a clear correlation between snow depth change and $\overline{\iDiv}_{2D}$ as we traverse an East-West-oriented ridgeline. Figure \ref{fig: Transect Compare}b shows a complementary transect (T2) oriented on a ridge orthogonal to T1. As we travel north along T2, $\Delta Hs$ and $\overline{\iDiv}_{2D}$ remain fairly constant, before reaching the more wind-exposed ridgeline ($\sim7747$ km North). Here, $\Delta Hs$ breaks and starts a steady decline, while $\overline{\iDiv}_{2D}$ also exhibits a shift towards smaller magnitude values. In fact, the start of the nearly linear decreases in $\Delta Hs$ and $\overline{\iDiv}_{2D}$ are offset by only 20 meters, suggesting a fairly precise connection between the flow physics predicted by $\overline{\iDiv}_{2D}$, and the cumulative effect of one week of storms on $\Delta Hs$. In both the T1 and T2 transects, ridgetop $\overline{\iDiv}_{2D}$ values begin to spread, which may be related to the strength of non-stationary turbulent processes when there is less equal-height terrain constraining the flow. Along T2, exactly at the turbulent ridgetop position where depth change becomes negative (7747.3 km), $\overline{\iDiv}_{2D}$ values shift back negative with large scatter.

\section{Discussion}
The Maxey-Riley (MR) equation was rigorously developed to account for all forces on a spherical inertial particle with small slip velocities in the Stokes drag regime \cite{Maxey1983}. We modified this equation to incorporate a particle Reynolds number dependent nonlinear drag which provided consistent dynamics against numerous particle benchmarks. We were able to significantly simplify snow velocity modeling with a reduced order model of the nonlinear MR system. This allows a simple calculation of temporally and spatially resolved snow particle velocities as a perturbation to the underlying fluid flow field. The ROM (\ref{eq: MR ROM}) replicates a variety of inertial particle behaviors across an unprecedented range of spatial scales, from individual particle slip velocities to kilometer scale accumulation patterns. Such a range of two-phase flow physical accuracy is currently unmatched with other snow modeling approaches.


At the meter-scale, the short timescale relationship between low wind speeds and particle suspension may at first appear counter-intuitive, but such behavior has also been measured in field studies. \citeA{Aksamit2017} found that near-surface ejections, bursts of slower than average streamwise wind, are often responsible for sustaining particles in transport and slowing settling. The long-term behavior of laterally converging particles may overcome much of the vertical divergence as falling particles that aggregate on vertically oriented flow structures eventually fall and accumulate on the ground. This appears to be the case especially for particles with larger inertia. This was experimentally verified by comparing temporally averaged near-surface inertial divergence with one week of preferential deposition measurements at an Arctic-alpine study site. Future work will dive deeper into the balance of timescales of attracting structures, lateral advection, and particle settling.

While our meter-scale cube experiment provides some insight into the short-term dynamics of snow particles and our ability to model them with reduced-order models, relying on brute force surface accumulation counting as a means to model accumulation patterns is problematic for multiple reasons. Integrating the full MR equations involves multiple differential equations with increased numerical sensitivity for small particles. This results in increased computational cost, as previously mentioned in Section \ref{Section: Sweep Loiter}. Even with perfectly performed integration, in practice one cannot guarantee that particles will accumulate in all regions, or that relative accumulation rates can be estimated everywhere. Depending on the coarseness of your ground accumulation grid, you could obtain zero information about regional accumulation rates, or be limited to only a very rough idea of spatial heterogeneity. In the present study, we counteracted this by integrating high density grids of particles at even greater computational cost. For example, for a single twenty-second accumulation window discussed above, 2.5 million particles required 50 times the computational time as a $600\times240$ grid of $\aLADtd$ on ten cores, and 4000 times as long as $\iDivtd$ on a single core. For applications in nature, obtaining accumulation patterns from large scale Lagrangian snow advection is unfeasible, making our efficient surface diagnostics even more advantageous.

In the context of hydrological modeling, this work provides a new and distinct avenue for snowdrift-permitting snowpack models. \citeA{Mott2018} classified existing models into two categories: vertically-integrated equilibrium flux models \cite<e.g.,>{Essery1999,Durand2005,Pomeroy2007,Liston2007}  and advection-diffusion driven models \cite<e.g.,>{Gauer1998,Lehning2008,Schneiderbauer2011,Vionnet2014,Vionnet2021}. Our approach makes progress towards a new class of model that falls into neither category. We avoid non-physical adaptations of log-law based transport profiles to terrain where no such wind profile exists, such as in the vertically-integrated models. Additionally, we more accurately represent the true inertial and nonlinear drag of snow than advection-diffusion models that treat snow concentration as a diffusive scalar. Our ROM also allows prediction of time-resolved snow transport coupled to non-stationary or coherent turbulent structures, which has not been a focus of snowpack modeling to date. Combined with a modeled precipitation rate, accumulation rates can be calculated directly using the mathematical definition of Lagrangian divergence (e.g., equation \ref{eq: Number Density Evo}).

The ROM incorporates non-linear drag and inertial effects into a particle velocity field that exhibits more realistic behavior than other approaches. For example, snow accumulation models that assume a fixed fall velocity, $\vvv_f$ \cite<e.g.,>{Lehning2004, Lehning2008}, modify the underlying wind flow with a constant offset, $\vvv=\uu - \vvv_f$. This means that $\mathrm{div}(\vvv)=\mathrm{div}_{2D}(\vvv)=0$ because spatial derivatives of velocity are indifferent to linear translations. Thus, a constant fall velocity model incorporates no inertial effects and allows no change in particle concentration while particles are in the air. Snow concentration changes (e.g., preferential deposition), therefore, can only arise as particles fall on the ground. This inherently neglects the true inertial forces and natural snowfall heterogeneity. Clearly, the transport inaccuracies are even greater for snow transport models that assume snow perfectly follows wind streamlines.

Our work opens the door to new approaches for various open problems in snow transport. The rate of sublimation of snow particles is intrinsically linked to the particle slip velocity. Until now, slip velocities could only be assumed constant, such as via an approximate terminal velocity, or after integrating particle velocity equations from an initial position and velocity guess. Using only physical parameters of snow particles, we can now calculate time-varying slip velocities directly from the underlying wind velocity and its derivatives. The snow velocity ROM can also be used to better understand momentum balances and transport rates in the saltation layer, immediately above the snow surface. Determining an asymptotic particle velocity profile in this region may help improve non-steady flow calculations, and avoid non-physical approximations in complex terrain.

\section{Conclusions}
The present work presents new reduced order models and accumulation diagnostics to predict spatially and temporally varying snow accumulation patterns in complex terrain. Inertial (snow) particle velocity fields were rigorously derived as a reduced order model (ROM) of a non-linear drag adaptation of the Maxey-Riley equation. The adapted MR equation was shown to faithfully recreate commonly-used empirically-derived hydrometeor terminal velocities. The ROM in turn accurately recreated multiple key aspects of falling snow particle dynamics. This includes accurate still-air terminal velocities for ranges of particle size and density, with well-defined boundaries and error estimates around those ranges. We also accurately recreate sweeping and loitering behavior of collections of particles in varying degrees of turbulence, at a significantly reduced computational cost (up to 1000-fold decrease) when compared with standard inertial particle modeling.

The ROM facilitated efficient computation of new accumulation diagnostics $\aLAD$ and $\aLADtd$, as well as their instantaneous counterparts $\iDiv$ and $\iDivtd$. Our accumulation diagnostics are derived directly from convergence and divergence in the inertial particle velocity fields. As such, they effectively describe the lateral and vertical separation of falling snow particles in turbulent flows, and outperformed wind speed descriptors. The resulting long-term accumulation was further verified with field observations of snow accumulation at an Arctic-alpine study site. Patterns in one week of snow depth change were accurately predicted by time-averaged $\iDivtd$ fields using only 500 seconds of simulation data. $\iDivtd$ significantly outperformed a common terrain-based accumulation diagnostic that was unable to predict accumulation on the correct side of several ridge lines.

Our new approach to coupled wind-snow modeling allows users to harness the high accuracy and fidelity of Lagrangian snow particle models with a thousand-fold increase in computational speed and a transparent (non-empirical) connection to the flow physics. Future work will move beyond predicting patterns in particle behavior, and connect the magnitude of our diagnostics to exact long-term accumulation rates. Further work to extend the ROM beyond our regime of small particle proxies is also underway.

Given the exponential growth of computing power and scientists' ability to run progressively higher resolution numerical weather and climate simulations, it is the perfect time to develop new snow transport models which better reflect other models' fidelity and physical accuracy. The ROM and diagnostics developed here provide an entirely new means to model and understand coupled wind-snow processes and minimize empiricism.

\section*{Acknowledgements}
The authors would like to recognize helpful discussions with Patrick Bardsley, George Haller, and Per Kristen Jakobsen regarding earlier iterations of this manuscript. We would also like to express appreciation to Karolina Haraldsson for insightful discussions about the JHTDB DNS. Furthermore, we would like to acknowledge some field equipment support from Dmitry Divine and the Norwegian Polar Institute.

\section*{Funding}
No external funding sources were used for the financing of this research.
\section*{Data Availability}
All data used in this manuscript is available at the cited public databases or was generated by open-source software following the methodology described in the body text.

\section*{Conflict of interest}
The authors declare no conflict of interest.

\section*{Author Contributions}
All authors contributed to the writing of the manuscript and interpretation of results. NA and AP developed the experiment design. NA and AEB developed the new theory and performed numerical simulations. NA and HH collected the field data and performed data analysis.
\section*{Graphical Abstract}
\begin{figure}[h!tbp]
    \RaggedLeft
    \includegraphics[width=0.25\linewidth]{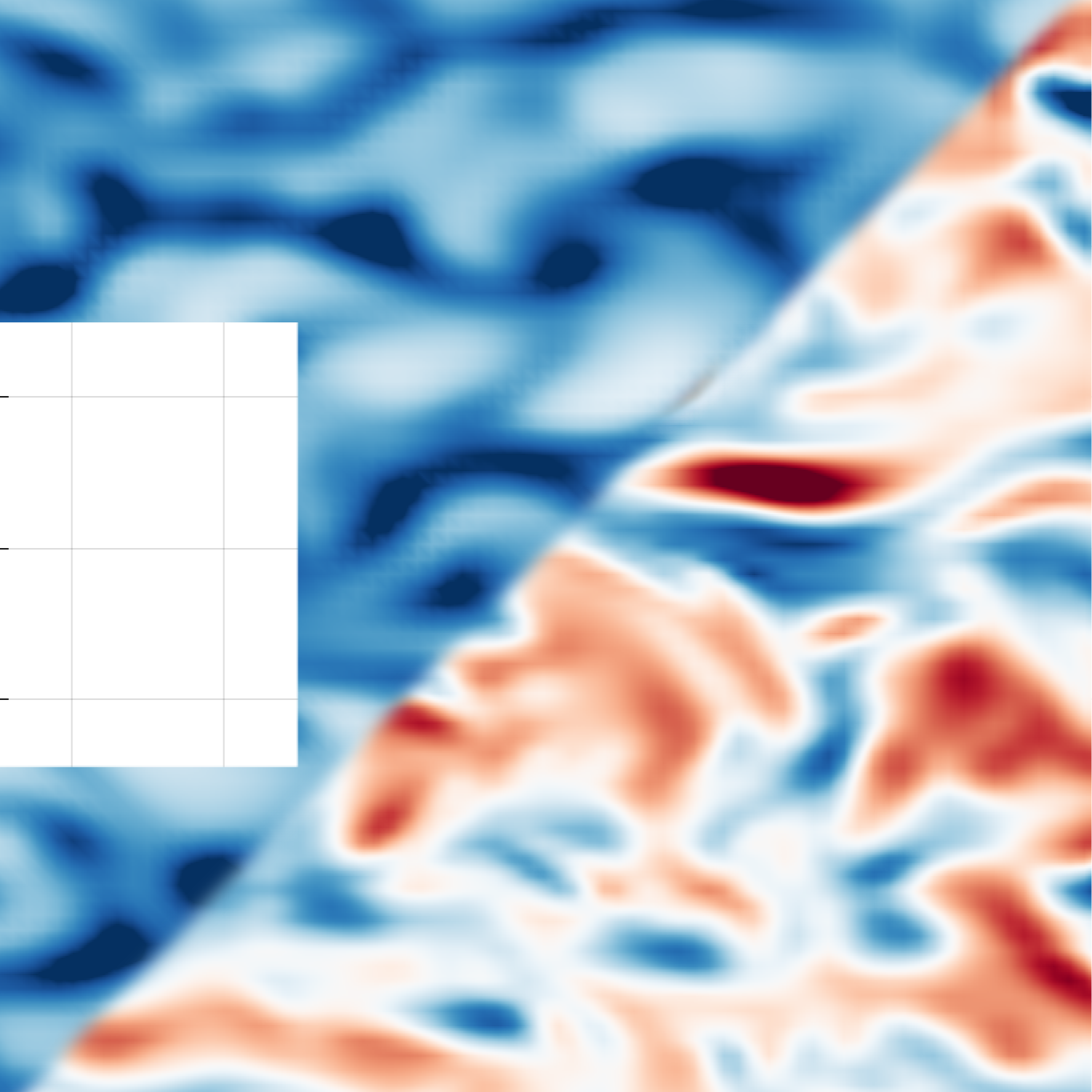}
    \caption*{Graphical Abstract: Our new reduced-order models of snow particle transport provide high-fidelity calculations of snow accumulation in turbulent flows at significantly reduced computational costs. Additional accumulation diagnostics from the reduced-order model predict complex patterns of particle concentration in turbulent boundary layers via coherent flow structures in both numerical simulations and field experiments.}
\end{figure}

%
\bibliography{references}

\begin{thebibliography}{}

\bibitem [\protect \citeauthoryear {%
Abraham%
}{%
Abraham%
}{%
{\protect \APACyear {1970}}%
}]{%
Abraham1970}
\APACinsertmetastar {%
Abraham1970}%
\begin{APACrefauthors}%
Abraham, F\BPBI F.%
\end{APACrefauthors}%
\unskip\
\newblock
\APACrefYearMonthDay{1970}{}{}.
\newblock
{\BBOQ}\APACrefatitle {{Functional dependence of drag coefficient of a sphere
  on reynolds number}} {{Functional dependence of drag coefficient of a sphere
  on reynolds number}}.{\BBCQ}
\newblock
\APACjournalVolNumPages{Physics of Fluids}{13}{8}{2194--2195}.
\newblock
\begin{APACrefDOI} \doi{10.1063/1.1693218} \end{APACrefDOI}
\PrintBackRefs{\CurrentBib}

\bibitem [\protect \citeauthoryear {%
Adam%
, Hamlet%
\BCBL {}\ \BBA {} Lettenmaier%
}{%
Adam%
\ \protect \BOthers {.}}{%
{\protect \APACyear {2009}}%
}]{%
Adam2009}
\APACinsertmetastar {%
Adam2009}%
\begin{APACrefauthors}%
Adam, J\BPBI C.%
, Hamlet, A\BPBI F.%
\BCBL {}\ \BBA {} Lettenmaier, D\BPBI P.%
\end{APACrefauthors}%
\unskip\
\newblock
\APACrefYearMonthDay{2009}{}{}.
\newblock
{\BBOQ}\APACrefatitle {{Implications of global climate change for snowmelt
  hydrology in the twenty-first century}} {{Implications of global climate
  change for snowmelt hydrology in the twenty-first century}}.{\BBCQ}
\newblock
\APACjournalVolNumPages{Hydrological Processes}{23}{}{962--972}.
\newblock
\begin{APACrefDOI} \doi{10.1002/hyp} \end{APACrefDOI}
\PrintBackRefs{\CurrentBib}

\bibitem [\protect \citeauthoryear {%
Aksamit%
\ \BBA {} Pomeroy%
}{%
Aksamit%
\ \BBA {} Pomeroy%
}{%
{\protect \APACyear {2017}}%
}]{%
Aksamit2017}
\APACinsertmetastar {%
Aksamit2017}%
\begin{APACrefauthors}%
Aksamit, N\BPBI O.%
\BCBT {}\ \BBA {} Pomeroy, J\BPBI W.%
\end{APACrefauthors}%
\unskip\
\newblock
\APACrefYearMonthDay{2017}{}{}.
\newblock
{\BBOQ}\APACrefatitle {{The Effect of Coherent Structures in the Atmospheric
  Surface Layer on Blowing-Snow Transport}} {{The Effect of Coherent Structures
  in the Atmospheric Surface Layer on Blowing-Snow Transport}}.{\BBCQ}
\newblock
\APACjournalVolNumPages{Boundary-Layer Meteorology}{167}{2}{1--23}.
\newblock
\begin{APACrefDOI} \doi{10.1007/s10546-017-0318-2} \end{APACrefDOI}
\PrintBackRefs{\CurrentBib}

\bibitem [\protect \citeauthoryear {%
Aksamit%
\ \BBA {} Pomeroy%
}{%
Aksamit%
\ \BBA {} Pomeroy%
}{%
{\protect \APACyear {2018}}%
}]{%
Aksamit2018}
\APACinsertmetastar {%
Aksamit2018}%
\begin{APACrefauthors}%
Aksamit, N\BPBI O.%
\BCBT {}\ \BBA {} Pomeroy, J\BPBI W.%
\end{APACrefauthors}%
\unskip\
\newblock
\APACrefYearMonthDay{2018}{}{}.
\newblock
{\BBOQ}\APACrefatitle {{Scale Interactions in Turbulence for Mountain Blowing
  Snow}} {{Scale Interactions in Turbulence for Mountain Blowing Snow}}.{\BBCQ}
\newblock
\APACjournalVolNumPages{Journal of Hydrometeorology}{19}{2}{305--320}.
\newblock
\begin{APACrefDOI} \doi{10.1175/JHM-D-17-0179.1} \end{APACrefDOI}
\PrintBackRefs{\CurrentBib}

\bibitem [\protect \citeauthoryear {%
Baker%
, Frankel%
, Mani%
\BCBL {}\ \BBA {} Coletti%
}{%
Baker%
\ \protect \BOthers {.}}{%
{\protect \APACyear {2017}}%
}]{%
Baker2017}
\APACinsertmetastar {%
Baker2017}%
\begin{APACrefauthors}%
Baker, L.%
, Frankel, A.%
, Mani, A.%
\BCBL {}\ \BBA {} Coletti, F.%
\end{APACrefauthors}%
\unskip\
\newblock
\APACrefYearMonthDay{2017}{}{}.
\newblock
{\BBOQ}\APACrefatitle {{Coherent clusters of inertial particles in homogeneous
  turbulence}} {{Coherent clusters of inertial particles in homogeneous
  turbulence}}.{\BBCQ}
\newblock
\APACjournalVolNumPages{Journal of Fluid Mechanics}{833}{}{364--398}.
\newblock
\begin{APACrefDOI} \doi{10.1017/jfm.2017.700} \end{APACrefDOI}
\PrintBackRefs{\CurrentBib}

\bibitem [\protect \citeauthoryear {%
B{\"{o}}hm%
}{%
B{\"{o}}hm%
}{%
{\protect \APACyear {1989}}%
}]{%
Bohm1989}
\APACinsertmetastar {%
Bohm1989}%
\begin{APACrefauthors}%
B{\"{o}}hm, H\BPBI P.%
\end{APACrefauthors}%
\unskip\
\newblock
\APACrefYearMonthDay{1989}{}{}.
\newblock
{\BBOQ}\APACrefatitle {{A General Equation for the Terminal Fall Speed of Solid
  Hydrometeors}} {{A General Equation for the Terminal Fall Speed of Solid
  Hydrometeors}}.{\BBCQ}
\newblock
\APACjournalVolNumPages{Journal of the Atmospheric
  Sciences}{46}{15}{2419--2427}.
\PrintBackRefs{\CurrentBib}

\bibitem [\protect \citeauthoryear {%
Brandt%
\ \BBA {} Coletti%
}{%
Brandt%
\ \BBA {} Coletti%
}{%
{\protect \APACyear {2022}}%
}]{%
Brandt2022}
\APACinsertmetastar {%
Brandt2022}%
\begin{APACrefauthors}%
Brandt, L.%
\BCBT {}\ \BBA {} Coletti, F.%
\end{APACrefauthors}%
\unskip\
\newblock
\APACrefYearMonthDay{2022}{}{}.
\newblock
{\BBOQ}\APACrefatitle {{Particle-Laden Turbulence : Progress and Perspectives}}
  {{Particle-Laden Turbulence : Progress and Perspectives}}.{\BBCQ}
\newblock
\APACjournalVolNumPages{Annual review of fluid mechanics}{54}{}{159--189}.
\PrintBackRefs{\CurrentBib}

\bibitem [\protect \citeauthoryear {%
Callaghan%
\ \protect \BOthers {.}}{%
Callaghan%
\ \protect \BOthers {.}}{%
{\protect \APACyear {2011}}%
}]{%
Callaghan2011}
\APACinsertmetastar {%
Callaghan2011}%
\begin{APACrefauthors}%
Callaghan, T\BPBI V.%
, Johansson, M.%
, Brown, R\BPBI D.%
, Groisman, P\BPBI Y.%
, Labba, N.%
, Radionov, V.%
\BDBL {}Wood, E\BPBI F.%
\end{APACrefauthors}%
\unskip\
\newblock
\APACrefYearMonthDay{2011}{}{}.
\newblock
{\BBOQ}\APACrefatitle {{Multiple effects of changes in arctic snow cover}}
  {{Multiple effects of changes in arctic snow cover}}.{\BBCQ}
\newblock
\APACjournalVolNumPages{Ambio}{40}{SUPPL. 1}{32--45}.
\newblock
\begin{APACrefDOI} \doi{10.1007/s13280-011-0213-x} \end{APACrefDOI}
\PrintBackRefs{\CurrentBib}

\bibitem [\protect \citeauthoryear {%
Cimino%
, Larsen%
, Johnston%
\BCBL {}\ \BBA {} Groff%
}{%
Cimino%
\ \protect \BOthers {.}}{%
{\protect \APACyear {2025}}%
}]{%
Cimino2025}
\APACinsertmetastar {%
Cimino2025}%
\begin{APACrefauthors}%
Cimino, M\BPBI A.%
, Larsen, G\BPBI D.%
, Johnston, D\BPBI W.%
\BCBL {}\ \BBA {} Groff, D\BPBI V.%
\end{APACrefauthors}%
\unskip\
\newblock
\APACrefYearMonthDay{2025}{}{}.
\newblock
{\BBOQ}\APACrefatitle {{Long ‑ term , landscape ‑ and wind ‑ driven snow
  conditions influence Ad{\'{e}}lie penguin colony extinctions}} {{Long ‑
  term , landscape ‑ and wind ‑ driven snow conditions influence
  Ad{\'{e}}lie penguin colony extinctions}}.{\BBCQ}
\newblock
\APACjournalVolNumPages{Landscape Ecology}{40:78}{}{}.
\newblock
\begin{APACrefDOI} \doi{10.1007/s10980-025-02088-y} \end{APACrefDOI}
\PrintBackRefs{\CurrentBib}

\bibitem [\protect \citeauthoryear {%
Comola%
, Giometto%
, Salesky%
, Parlange%
\BCBL {}\ \BBA {} Lehning%
}{%
Comola%
\ \protect \BOthers {.}}{%
{\protect \APACyear {2019}}%
}]{%
Comola2019}
\APACinsertmetastar {%
Comola2019}%
\begin{APACrefauthors}%
Comola, F.%
, Giometto, M\BPBI G.%
, Salesky, S\BPBI T.%
, Parlange, M\BPBI B.%
\BCBL {}\ \BBA {} Lehning, M.%
\end{APACrefauthors}%
\unskip\
\newblock
\APACrefYearMonthDay{2019}{}{}.
\newblock
{\BBOQ}\APACrefatitle {{Preferential Deposition of Snow and Dust Over Hills:
  Governing Processes and Relevant Scales}} {{Preferential Deposition of Snow
  and Dust Over Hills: Governing Processes and Relevant Scales}}.{\BBCQ}
\newblock
\APACjournalVolNumPages{Journal of Geophysical Research:
  Atmospheres}{124}{14}{7951--7974}.
\newblock
\begin{APACrefDOI} \doi{10.1029/2018JD029614} \end{APACrefDOI}
\PrintBackRefs{\CurrentBib}

\bibitem [\protect \citeauthoryear {%
Dadic%
, Mott%
, Lehning%
\BCBL {}\ \BBA {} Burlando%
}{%
Dadic%
\ \protect \BOthers {.}}{%
{\protect \APACyear {2010}}%
}]{%
Dadic2010}
\APACinsertmetastar {%
Dadic2010}%
\begin{APACrefauthors}%
Dadic, R.%
, Mott, R.%
, Lehning, M.%
\BCBL {}\ \BBA {} Burlando, P.%
\end{APACrefauthors}%
\unskip\
\newblock
\APACrefYearMonthDay{2010}{}{}.
\newblock
{\BBOQ}\APACrefatitle {{Wind influence on snow depth distribution and
  accumulation over glaciers}} {{Wind influence on snow depth distribution and
  accumulation over glaciers}}.{\BBCQ}
\newblock
\APACjournalVolNumPages{Journal of Geophysical Research: Earth
  Surface}{115}{1}{1--8}.
\newblock
\begin{APACrefDOI} \doi{10.1029/2009JF001261} \end{APACrefDOI}
\PrintBackRefs{\CurrentBib}

\bibitem [\protect \citeauthoryear {%
Davies%
}{%
Davies%
}{%
{\protect \APACyear {1945}}%
}]{%
Davies1945}
\APACinsertmetastar {%
Davies1945}%
\begin{APACrefauthors}%
Davies, C\BPBI N.%
\end{APACrefauthors}%
\unskip\
\newblock
\APACrefYearMonthDay{1945}{}{}.
\newblock
{\BBOQ}\APACrefatitle {{Definitive equations for the fluid resistance of
  spheres}} {{Definitive equations for the fluid resistance of
  spheres}}.{\BBCQ}
\newblock
\APACjournalVolNumPages{Proceedings of the Physical Society}{57}{4}{259--270}.
\newblock
\begin{APACrefDOI} \doi{10.1088/0959-5309/57/4/301} \end{APACrefDOI}
\PrintBackRefs{\CurrentBib}

\bibitem [\protect \citeauthoryear {%
Durand%
, Guyomarc'h%
, M{\'{e}}rindol%
\BCBL {}\ \BBA {} Corripio%
}{%
Durand%
\ \protect \BOthers {.}}{%
{\protect \APACyear {2005}}%
}]{%
Durand2005}
\APACinsertmetastar {%
Durand2005}%
\begin{APACrefauthors}%
Durand, Y.%
, Guyomarc'h, G.%
, M{\'{e}}rindol, L.%
\BCBL {}\ \BBA {} Corripio, J\BPBI G.%
\end{APACrefauthors}%
\unskip\
\newblock
\APACrefYearMonthDay{2005}{}{}.
\newblock
{\BBOQ}\APACrefatitle {{Improvement of a numerical snow drift model and field
  validation}} {{Improvement of a numerical snow drift model and field
  validation}}.{\BBCQ}
\newblock
\APACjournalVolNumPages{Cold Regions Science and Technology}{43}{1-2}{93--103}.
\newblock
\begin{APACrefDOI} \doi{10.1016/j.coldregions.2005.05.008} \end{APACrefDOI}
\PrintBackRefs{\CurrentBib}

\bibitem [\protect \citeauthoryear {%
Essery%
, Li%
\BCBL {}\ \BBA {} Pomeroy%
}{%
Essery%
\ \protect \BOthers {.}}{%
{\protect \APACyear {1999}}%
}]{%
Essery1999}
\APACinsertmetastar {%
Essery1999}%
\begin{APACrefauthors}%
Essery, R.%
, Li, L.%
\BCBL {}\ \BBA {} Pomeroy, J.%
\end{APACrefauthors}%
\unskip\
\newblock
\APACrefYearMonthDay{1999}{}{}.
\newblock
{\BBOQ}\APACrefatitle {{A distributed model of blowing snow over complex
  terrain}} {{A distributed model of blowing snow over complex
  terrain}}.{\BBCQ}
\newblock
\APACjournalVolNumPages{Hydrological Processes}{13}{}{2423--2438}.
\newblock
\begin{APACrefDOI}
  \doi{10.1002/(SICI)1099-1085(199910)13:14/15<2423::AID-HYP853>3.0.CO;2-U}
  \end{APACrefDOI}
\PrintBackRefs{\CurrentBib}

\bibitem [\protect \citeauthoryear {%
Fenichel%
}{%
Fenichel%
}{%
{\protect \APACyear {1979}}%
}]{%
Fenichel1979}
\APACinsertmetastar {%
Fenichel1979}%
\begin{APACrefauthors}%
Fenichel, N.%
\end{APACrefauthors}%
\unskip\
\newblock
\APACrefYearMonthDay{1979}{}{}.
\newblock
{\BBOQ}\APACrefatitle {{Geometric singular perturbation theory for ordinary
  differential equations}} {{Geometric singular perturbation theory for
  ordinary differential equations}}.{\BBCQ}
\newblock
\APACjournalVolNumPages{Journal of Differential Equations}{31}{1}{53--98}.
\newblock
\begin{APACrefDOI} \doi{10.1016/0022-0396(79)90152-9} \end{APACrefDOI}
\PrintBackRefs{\CurrentBib}

\bibitem [\protect \citeauthoryear {%
Gauer%
}{%
Gauer%
}{%
{\protect \APACyear {1998}}%
}]{%
Gauer1998}
\APACinsertmetastar {%
Gauer1998}%
\begin{APACrefauthors}%
Gauer, P.%
\end{APACrefauthors}%
\unskip\
\newblock
\APACrefYearMonthDay{1998}{}{}.
\newblock
{\BBOQ}\APACrefatitle {{Blowing and drifting snow in alpine terrain: A
  physically based numerical model and related field measurements}} {{Blowing
  and drifting snow in alpine terrain: A physically based numerical model and
  related field measurements}}.{\BBCQ}
\newblock
\APACjournalVolNumPages{Annals of Glaciology}{26}{}{174--178}.
\PrintBackRefs{\CurrentBib}

\bibitem [\protect \citeauthoryear {%
Good%
\ \protect \BOthers {.}}{%
Good%
\ \protect \BOthers {.}}{%
{\protect \APACyear {2014}}%
}]{%
Good2014}
\APACinsertmetastar {%
Good2014}%
\begin{APACrefauthors}%
Good, G\BPBI H.%
, Ireland, P\BPBI J.%
, Bewley, G\BPBI P.%
, Bodenschatz, E.%
, Collins, L\BPBI R.%
\BCBL {}\ \BBA {} Warhaft, Z.%
\end{APACrefauthors}%
\unskip\
\newblock
\APACrefYearMonthDay{2014}{}{}.
\newblock
{\BBOQ}\APACrefatitle {{Settling regimes of inertial particles in isotropic
  turbulence}} {{Settling regimes of inertial particles in isotropic
  turbulence}}.{\BBCQ}
\newblock
\APACjournalVolNumPages{Journal of Fluid Mechanics}{759}{}{R3}.
\newblock
\begin{APACrefDOI} \doi{10.1017/jfm.2014.602} \end{APACrefDOI}
\PrintBackRefs{\CurrentBib}

\bibitem [\protect \citeauthoryear {%
Gordon%
\ \BBA {} Taylor%
}{%
Gordon%
\ \BBA {} Taylor%
}{%
{\protect \APACyear {2009}}%
}]{%
Gordon2009}
\APACinsertmetastar {%
Gordon2009}%
\begin{APACrefauthors}%
Gordon, M.%
\BCBT {}\ \BBA {} Taylor, P\BPBI A.%
\end{APACrefauthors}%
\unskip\
\newblock
\APACrefYearMonthDay{2009}{}{}.
\newblock
{\BBOQ}\APACrefatitle {{Measurements of blowing snow, Part I: Particle shape,
  size distribution, velocity, and number flux at Churchill, Manitoba, Canada}}
  {{Measurements of blowing snow, Part I: Particle shape, size distribution,
  velocity, and number flux at Churchill, Manitoba, Canada}}.{\BBCQ}
\newblock
\APACjournalVolNumPages{Cold Regions Science and Technology}{55}{1}{63--74}.
\newblock
\begin{APACrefDOI} \doi{10.1016/j.coldregions.2008.05.001} \end{APACrefDOI}
\PrintBackRefs{\CurrentBib}

\bibitem [\protect \citeauthoryear {%
{Groot Zwaaftink}%
\ \protect \BOthers {.}}{%
{Groot Zwaaftink}%
\ \protect \BOthers {.}}{%
{\protect \APACyear {2014}}%
}]{%
GrootZwaaftink2014}
\APACinsertmetastar {%
GrootZwaaftink2014}%
\begin{APACrefauthors}%
{Groot Zwaaftink}, C\BPBI D.%
, Diebold, M.%
, Horender, S.%
, Overney, J.%
, Lieberherr, G.%
, Parlange, M\BPBI B.%
\BCBL {}\ \BBA {} Lehning, M.%
\end{APACrefauthors}%
\unskip\
\newblock
\APACrefYearMonthDay{2014}{}{}.
\newblock
{\BBOQ}\APACrefatitle {{Modelling Small-Scale Drifting Snow with a Lagrangian
  Stochastic Model Based on Large-Eddy Simulations}} {{Modelling Small-Scale
  Drifting Snow with a Lagrangian Stochastic Model Based on Large-Eddy
  Simulations}}.{\BBCQ}
\newblock
\APACjournalVolNumPages{Boundary-Layer Meteorology}{153}{1}{117--139}.
\newblock
\begin{APACrefDOI} \doi{10.1007/s10546-014-9934-2} \end{APACrefDOI}
\PrintBackRefs{\CurrentBib}

\bibitem [\protect \citeauthoryear {%
Haller%
}{%
Haller%
}{%
{\protect \APACyear {2023}}%
}]{%
Haller}
\APACinsertmetastar {%
Haller}%
\begin{APACrefauthors}%
Haller, G.%
\end{APACrefauthors}%
\unskip\
\newblock
\APACrefYear{2023}.
\newblock
\APACrefbtitle {{Transport Barriers in Flow Data: Advective, Diffusive,
  Stochastic and Active Methods}} {{Transport Barriers in Flow Data: Advective,
  Diffusive, Stochastic and Active Methods}}.
\newblock
\APACaddressPublisher{Cambridge, UK}{Cambridge University Press}.
\PrintBackRefs{\CurrentBib}

\bibitem [\protect \citeauthoryear {%
Haller%
\ \BBA {} Sapsis%
}{%
Haller%
\ \BBA {} Sapsis%
}{%
{\protect \APACyear {2008}}%
}]{%
Haller2008}
\APACinsertmetastar {%
Haller2008}%
\begin{APACrefauthors}%
Haller, G.%
\BCBT {}\ \BBA {} Sapsis, T.%
\end{APACrefauthors}%
\unskip\
\newblock
\APACrefYearMonthDay{2008}{}{}.
\newblock
{\BBOQ}\APACrefatitle {{Where do inertial particles go in fluid flows?}}
  {{Where do inertial particles go in fluid flows?}}{\BBCQ}
\newblock
\APACjournalVolNumPages{Physica D: Nonlinear Phenomena}{237}{5}{573--583}.
\newblock
\begin{APACrefDOI} \doi{10.1016/j.physd.2007.09.027} \end{APACrefDOI}
\PrintBackRefs{\CurrentBib}

\bibitem [\protect \citeauthoryear {%
Hames%
, Jafari%
, Köhler%
, Haas%
\BCBL {}\ \BBA {} Lehning%
}{%
Hames%
\ \protect \BOthers {.}}{%
{\protect \APACyear {2025}}%
}]{%
Hames2025}
\APACinsertmetastar {%
Hames2025}%
\begin{APACrefauthors}%
Hames, O.%
, Jafari, M.%
, Köhler, P.%
, Haas, C.%
\BCBL {}\ \BBA {} Lehning, M.%
\end{APACrefauthors}%
\unskip\
\newblock
\APACrefYearMonthDay{2025}{}{}.
\newblock
{\BBOQ}\APACrefatitle {Governing Processes of Structure-Borne Snowdrifts: A
  Case Study at Neumayer Station III} {Governing processes of structure-borne
  snowdrifts: A case study at neumayer station iii}.{\BBCQ}
\newblock
\APACjournalVolNumPages{Journal of Geophysical Research: Earth
  Surface}{130}{3}{}.
\newblock
\begin{APACrefDOI} \doi{10.1029/2024JF008180} \end{APACrefDOI}
\PrintBackRefs{\CurrentBib}

\bibitem [\protect \citeauthoryear {%
Hames%
\ \protect \BOthers {.}}{%
Hames%
\ \protect \BOthers {.}}{%
{\protect \APACyear {2022}}%
}]{%
Hames2022}
\APACinsertmetastar {%
Hames2022}%
\begin{APACrefauthors}%
Hames, O.%
, Jafari, M.%
, Wagner, D\BPBI N.%
, Raphael, I.%
, Clemens-Sewall, D.%
, Polashenski, C.%
\BDBL {}Lehning, M.%
\end{APACrefauthors}%
\unskip\
\newblock
\APACrefYearMonthDay{2022}{}{}.
\newblock
{\BBOQ}\APACrefatitle {{Modeling the small-scale deposition of snow onto
  structured Arctic sea ice during a MOSAiC storm using snowBedFoam 1.0.}}
  {{Modeling the small-scale deposition of snow onto structured Arctic sea ice
  during a MOSAiC storm using snowBedFoam 1.0.}}{\BBCQ}
\newblock
\APACjournalVolNumPages{Geoscientific Model Development}{15}{16}{6429--6449}.
\newblock
\begin{APACrefDOI} \doi{10.5194/gmd-15-6429-2022} \end{APACrefDOI}
\PrintBackRefs{\CurrentBib}

\bibitem [\protect \citeauthoryear {%
Hersbach%
\ \protect \BOthers {.}}{%
Hersbach%
\ \protect \BOthers {.}}{%
{\protect \APACyear {2023}}%
}]{%
hersbach2023era5}
\APACinsertmetastar {%
hersbach2023era5}%
\begin{APACrefauthors}%
Hersbach, H.%
, Bell, B.%
, Berrisford, P.%
, Biavati, G.%
, Horányi, A.%
, Muñoz~Sabater, J.%
\BDBL {}Thépaut, J\BHBI N.%
\end{APACrefauthors}%
\unskip\
\newblock
\APACrefYearMonthDay{2023}{}{}.
\newblock
\APACrefbtitle {ERA5 hourly data on single levels from 1940 to present.} {Era5
  hourly data on single levels from 1940 to present.}
\newblock
\APACaddressPublisher{}{Copernicus Climate Change Service (C3S) Climate Data
  Store (CDS)}.
\newblock
\APACrefnote{Accessed on 01-Feb-2025}
\newblock
\begin{APACrefDOI} \doi{10.24381/cds.adbb2d47} \end{APACrefDOI}
\PrintBackRefs{\CurrentBib}

\bibitem [\protect \citeauthoryear {%
Heymsfield%
\ \BBA {} Westbrook%
}{%
Heymsfield%
\ \BBA {} Westbrook%
}{%
{\protect \APACyear {2010}}%
}]{%
Heymsfield2010}
\APACinsertmetastar {%
Heymsfield2010}%
\begin{APACrefauthors}%
Heymsfield, A\BPBI J.%
\BCBT {}\ \BBA {} Westbrook, C\BPBI D.%
\end{APACrefauthors}%
\unskip\
\newblock
\APACrefYearMonthDay{2010}{}{}.
\newblock
{\BBOQ}\APACrefatitle {{Advances in the estimation of ice particle fall speeds
  using laboratory and field measurements}} {{Advances in the estimation of ice
  particle fall speeds using laboratory and field measurements}}.{\BBCQ}
\newblock
\APACjournalVolNumPages{Journal of the Atmospheric
  Sciences}{67}{8}{2469--2482}.
\newblock
\begin{APACrefDOI} \doi{10.1175/2010JAS3379.1} \end{APACrefDOI}
\PrintBackRefs{\CurrentBib}

\bibitem [\protect \citeauthoryear {%
Hock%
\ \protect \BOthers {.}}{%
Hock%
\ \protect \BOthers {.}}{%
{\protect \APACyear {2019}}%
}]{%
Hock2019HighMountain}
\APACinsertmetastar {%
Hock2019HighMountain}%
\begin{APACrefauthors}%
Hock, R.%
, Bacchiocchi, D.%
, Pelto, M.%
, Davies, B.%
, Rowan, A.%
, Kääb, A.%
\BCBL {}\ \BBA {} et al.%
\end{APACrefauthors}%
\unskip\
\newblock
\APACrefYearMonthDay{2019}{}{}.
\newblock
{\BBOQ}\APACrefatitle {High Mountain Areas} {High mountain areas}.{\BBCQ}
\newblock
\BIn{} H\BHBI O.~Pörtner\ \BOthers {.}\ (\BEDS), \APACrefbtitle {IPCC Special
  Report on the Ocean and Cryosphere in a Changing Climate} {Ipcc special
  report on the ocean and cryosphere in a changing climate}\ (\BPGS\ 131--202).
\newblock
\APACaddressPublisher{}{Cambridge University Press}.
\newblock
\begin{APACrefDOI} \doi{10.1017/9781009157964.004} \end{APACrefDOI}
\PrintBackRefs{\CurrentBib}

\bibitem [\protect \citeauthoryear {%
Ishizaka%
}{%
Ishizaka%
}{%
{\protect \APACyear {1993}}%
}]{%
Ishizaka1993}
\APACinsertmetastar {%
Ishizaka1993}%
\begin{APACrefauthors}%
Ishizaka, M.%
\end{APACrefauthors}%
\unskip\
\newblock
\APACrefYearMonthDay{1993}{}{}.
\newblock
{\BBOQ}\APACrefatitle {{An accurate measurement of densities of snowflakes
  using 3-D microphotographs}} {{An accurate measurement of densities of
  snowflakes using 3-D microphotographs}}.{\BBCQ}
\newblock
\APACjournalVolNumPages{Annals of Glaciology}{18}{}{92--96}.
\newblock
\begin{APACrefDOI} \doi{10.3189/s0260305500011319} \end{APACrefDOI}
\PrintBackRefs{\CurrentBib}

\bibitem [\protect \citeauthoryear {%
Ishizaka%
\ \protect \BOthers {.}}{%
Ishizaka%
\ \protect \BOthers {.}}{%
{\protect \APACyear {2016}}%
}]{%
Ishizaka2016}
\APACinsertmetastar {%
Ishizaka2016}%
\begin{APACrefauthors}%
Ishizaka, M.%
, Motoyoshi, H.%
, Yamaguchi, S.%
, Nakai, S.%
, Shiina, T.%
\BCBL {}\ \BBA {} Muramoto, K\BPBI I.%
\end{APACrefauthors}%
\unskip\
\newblock
\APACrefYearMonthDay{2016}{}{}.
\newblock
{\BBOQ}\APACrefatitle {{Relationships between snowfall density and solid
  hydrometeors, based on measured size and fall speed, for snowpack modeling
  applications}} {{Relationships between snowfall density and solid
  hydrometeors, based on measured size and fall speed, for snowpack modeling
  applications}}.{\BBCQ}
\newblock
\APACjournalVolNumPages{Cryosphere}{10}{6}{2831--2845}.
\newblock
\begin{APACrefDOI} \doi{10.5194/tc-10-2831-2016} \end{APACrefDOI}
\PrintBackRefs{\CurrentBib}

\bibitem [\protect \citeauthoryear {%
{Kartverket}%
}{%
{Kartverket}%
}{%
{\protect \APACyear {2014}}%
}]{%
Kartverket2014}
\APACinsertmetastar {%
Kartverket2014}%
\begin{APACrefauthors}%
{Kartverket}.%
\end{APACrefauthors}%
\unskip\
\newblock
\APACrefYearMonthDay{2014}{}{}.
\newblock
\APACrefbtitle {{Forprosjekt “Nasjonal, detaljert høydemodell”}}
  {{Forprosjekt “Nasjonal, detaljert høydemodell”}}\
  \APACbVolEdTR{}{\BTR{}}.
\newblock
\APACaddressInstitution{}{Norwegian Mapping Authority}.
\newblock
\begin{APACrefURL} \url{https://hoydedata.no/LaserInnsyn/} \end{APACrefURL}
\newblock
\APACrefnote{Last accessed: 5 August 2024}
\PrintBackRefs{\CurrentBib}

\bibitem [\protect \citeauthoryear {%
Lehning%
\ \protect \BOthers {.}}{%
Lehning%
\ \protect \BOthers {.}}{%
{\protect \APACyear {2004}}%
}]{%
Lehning2004}
\APACinsertmetastar {%
Lehning2004}%
\begin{APACrefauthors}%
Lehning, M.%
, Bartelt, P.%
, Bethke, S.%
, Fierz, C.%
, Gustafsson, D.%
, Landl, B.%
\BDBL {}St{\"a}hli, M.%
\end{APACrefauthors}%
\unskip\
\newblock
\APACrefYearMonthDay{2004}{}{}.
\newblock
{\BBOQ}\APACrefatitle {Review of SNOWPACK and ALPINE3D applications} {Review of
  snowpack and alpine3d applications}.{\BBCQ}
\newblock
\BIn{} P.~Bartelt, R.~Sack, A.~Sato, E.~Adams\BCBL {}\ \BBA {} M.~Christen\
  (\BEDS), \APACrefbtitle {Snow Engineering} {Snow engineering}\ (\BPGS\
  299--307).
\newblock
\APACaddressPublisher{The Netherlands}{Balkema}.
\PrintBackRefs{\CurrentBib}

\bibitem [\protect \citeauthoryear {%
Lehning%
\ \BBA {} Fierz%
}{%
Lehning%
\ \BBA {} Fierz%
}{%
{\protect \APACyear {2008}}%
}]{%
Lehning2008}
\APACinsertmetastar {%
Lehning2008}%
\begin{APACrefauthors}%
Lehning, M.%
\BCBT {}\ \BBA {} Fierz, C.%
\end{APACrefauthors}%
\unskip\
\newblock
\APACrefYearMonthDay{2008}{}{}.
\newblock
{\BBOQ}\APACrefatitle {{Assessment of snow transport in avalanche terrain}}
  {{Assessment of snow transport in avalanche terrain}}.{\BBCQ}
\newblock
\APACjournalVolNumPages{Cold Regions Science and
  Technology}{51}{2-3}{240--252}.
\newblock
\begin{APACrefDOI} \doi{10.1016/j.coldregions.2007.05.012} \end{APACrefDOI}
\PrintBackRefs{\CurrentBib}

\bibitem [\protect \citeauthoryear {%
J.~Li%
, Guala%
\BCBL {}\ \BBA {} Hong%
}{%
J.~Li%
\ \protect \BOthers {.}}{%
{\protect \APACyear {2023}}%
}]{%
Li2023}
\APACinsertmetastar {%
Li2023}%
\begin{APACrefauthors}%
Li, J.%
, Guala, M.%
\BCBL {}\ \BBA {} Hong, J.%
\end{APACrefauthors}%
\unskip\
\newblock
\APACrefYearMonthDay{2023}{}{}.
\newblock
{\BBOQ}\APACrefatitle {{Snow Particle Analyzer for Simultaneous Measurements of
  Snow Density and Morphology}} {{Snow Particle Analyzer for Simultaneous
  Measurements of Snow Density and Morphology}}.{\BBCQ}
\newblock
\APACjournalVolNumPages{Journal of Geophysical Research:
  Atmospheres}{128}{16}{}.
\newblock
\begin{APACrefDOI} \doi{10.1029/2023JD038987} \end{APACrefDOI}
\PrintBackRefs{\CurrentBib}

\bibitem [\protect \citeauthoryear {%
J.~Li%
, Guala%
\BCBL {}\ \BBA {} Hong%
}{%
J.~Li%
\ \protect \BOthers {.}}{%
{\protect \APACyear {2024}}%
}]{%
Li2024}
\APACinsertmetastar {%
Li2024}%
\begin{APACrefauthors}%
Li, J.%
, Guala, M.%
\BCBL {}\ \BBA {} Hong, J.%
\end{APACrefauthors}%
\unskip\
\newblock
\APACrefYearMonthDay{2024}{}{}.
\newblock
{\BBOQ}\APACrefatitle {{Field investigation of 3D snow settling dynamics under
  weak atmospheric turbulence}} {{Field investigation of 3D snow settling
  dynamics under weak atmospheric turbulence}}.{\BBCQ}
\newblock
\APACjournalVolNumPages{Journal of Fluid Mechanics}{997}{}{A33}.
\newblock
\begin{APACrefDOI} \doi{10.1017/jfm.2024.601} \end{APACrefDOI}
\PrintBackRefs{\CurrentBib}

\bibitem [\protect \citeauthoryear {%
Y.~Li%
\ \protect \BOthers {.}}{%
Y.~Li%
\ \protect \BOthers {.}}{%
{\protect \APACyear {2008}}%
}]{%
Li2008}
\APACinsertmetastar {%
Li2008}%
\begin{APACrefauthors}%
Li, Y.%
, Perlman, E.%
, Wan, M.%
, Yang, Y.%
, Meneveau, C.%
, Burns, R.%
\BDBL {}Eyink, G.%
\end{APACrefauthors}%
\unskip\
\newblock
\APACrefYearMonthDay{2008}{}{}.
\newblock
{\BBOQ}\APACrefatitle {A public turbulence database cluster and applications to
  study Lagrangian evolution of velocity increments in turbulence} {A public
  turbulence database cluster and applications to study lagrangian evolution of
  velocity increments in turbulence}.{\BBCQ}
\newblock
\APACjournalVolNumPages{Journal of Turbulence}{9}{31}{}.
\newblock
\begin{APACrefDOI} \doi{10.1080/14685240802376389} \end{APACrefDOI}
\PrintBackRefs{\CurrentBib}

\bibitem [\protect \citeauthoryear {%
Liston%
\ \protect \BOthers {.}}{%
Liston%
\ \protect \BOthers {.}}{%
{\protect \APACyear {2007}}%
}]{%
Liston2007}
\APACinsertmetastar {%
Liston2007}%
\begin{APACrefauthors}%
Liston, G\BPBI E.%
, Haehnel, R\BPBI B.%
, Sturm, M.%
, Hiemstra, C\BPBI A.%
, Berezovskaya, S.%
\BCBL {}\ \BBA {} Tabler, R\BPBI D.%
\end{APACrefauthors}%
\unskip\
\newblock
\APACrefYearMonthDay{2007}{}{}.
\newblock
{\BBOQ}\APACrefatitle {{Simulating complex snow distributions in windy
  environments using SnowTran-3D}} {{Simulating complex snow distributions in
  windy environments using SnowTran-3D}}.{\BBCQ}
\newblock
\APACjournalVolNumPages{Journal of Glaciology}{53}{181}{241--256}.
\newblock
\begin{APACrefDOI} \doi{10.3189/172756507782202865} \end{APACrefDOI}
\PrintBackRefs{\CurrentBib}

\bibitem [\protect \citeauthoryear {%
Lovecchio%
, Marchioli%
\BCBL {}\ \BBA {} Soldati%
}{%
Lovecchio%
\ \protect \BOthers {.}}{%
{\protect \APACyear {2013}}%
}]{%
Lovecchio2013}
\APACinsertmetastar {%
Lovecchio2013}%
\begin{APACrefauthors}%
Lovecchio, S.%
, Marchioli, C.%
\BCBL {}\ \BBA {} Soldati, A.%
\end{APACrefauthors}%
\unskip\
\newblock
\APACrefYearMonthDay{2013}{}{}.
\newblock
{\BBOQ}\APACrefatitle {{Time persistence of floating-particle clusters in
  free-surface turbulence}} {{Time persistence of floating-particle clusters in
  free-surface turbulence}}.{\BBCQ}
\newblock
\APACjournalVolNumPages{Physical Review E - Statistical, Nonlinear, and Soft
  Matter Physics}{88}{3}{1--6}.
\newblock
\begin{APACrefDOI} \doi{10.1103/PhysRevE.88.033003} \end{APACrefDOI}
\PrintBackRefs{\CurrentBib}

\bibitem [\protect \citeauthoryear {%
Lussana%
, Tveito%
, Dobler%
\BCBL {}\ \BBA {} Tunheim%
}{%
Lussana%
\ \protect \BOthers {.}}{%
{\protect \APACyear {2019}}%
}]{%
Lussana2019}
\APACinsertmetastar {%
Lussana2019}%
\begin{APACrefauthors}%
Lussana, C.%
, Tveito, O\BPBI E.%
, Dobler, A.%
\BCBL {}\ \BBA {} Tunheim, K.%
\end{APACrefauthors}%
\unskip\
\newblock
\APACrefYearMonthDay{2019}{}{}.
\newblock
{\BBOQ}\APACrefatitle {seNorge\_2018, daily precipitation, and temperature
  datasets over Norway} {senorge\_2018, daily precipitation, and temperature
  datasets over norway}.{\BBCQ}
\newblock
\APACjournalVolNumPages{Earth System Science Data}{11}{4}{1531--1551}.
\newblock
\begin{APACrefDOI} \doi{10.5194/essd-11-1531-2019} \end{APACrefDOI}
\PrintBackRefs{\CurrentBib}

\bibitem [\protect \citeauthoryear {%
Maronga%
\ \protect \BOthers {.}}{%
Maronga%
\ \protect \BOthers {.}}{%
{\protect \APACyear {2020}}%
}]{%
Maronga2020}
\APACinsertmetastar {%
Maronga2020}%
\begin{APACrefauthors}%
Maronga, B.%
, Banzhaf, S.%
, Burmeister, C.%
, Esch, T.%
, Forkel, R.%
, Fr{\"{o}}hlich, D.%
\BDBL {}Raasch, S.%
\end{APACrefauthors}%
\unskip\
\newblock
\APACrefYearMonthDay{2020}{}{}.
\newblock
{\BBOQ}\APACrefatitle {{Overview of the PALM model system 6.0}} {{Overview of
  the PALM model system 6.0}}.{\BBCQ}
\newblock
\APACjournalVolNumPages{Geoscientific Model Development}{13}{3}{1335--1372}.
\newblock
\begin{APACrefDOI} \doi{10.5194/gmd-13-1335-2020} \end{APACrefDOI}
\PrintBackRefs{\CurrentBib}

\bibitem [\protect \citeauthoryear {%
Marsh%
, Pomeroy%
, Spiteri%
\BCBL {}\ \BBA {} Wheater%
}{%
Marsh%
\ \protect \BOthers {.}}{%
{\protect \APACyear {2020}}%
}]{%
Marsh2020}
\APACinsertmetastar {%
Marsh2020}%
\begin{APACrefauthors}%
Marsh, C\BPBI B.%
, Pomeroy, J\BPBI W.%
, Spiteri, R\BPBI J.%
\BCBL {}\ \BBA {} Wheater, H\BPBI S.%
\end{APACrefauthors}%
\unskip\
\newblock
\APACrefYearMonthDay{2020}{}{}.
\newblock
{\BBOQ}\APACrefatitle {{A Finite Volume Blowing Snow Model for Use With
  Variable Resolution Meshes}} {{A Finite Volume Blowing Snow Model for Use
  With Variable Resolution Meshes}}.{\BBCQ}
\newblock
\APACjournalVolNumPages{Water Resources Research}{56}{2}{1--28}.
\newblock
\begin{APACrefDOI} \doi{10.1029/2019WR025307} \end{APACrefDOI}
\PrintBackRefs{\CurrentBib}

\bibitem [\protect \citeauthoryear {%
Maxey%
\ \BBA {} Riley%
}{%
Maxey%
\ \BBA {} Riley%
}{%
{\protect \APACyear {1983}}%
}]{%
Maxey1983}
\APACinsertmetastar {%
Maxey1983}%
\begin{APACrefauthors}%
Maxey, M.%
\BCBT {}\ \BBA {} Riley, J.%
\end{APACrefauthors}%
\unskip\
\newblock
\APACrefYearMonthDay{1983}{}{}.
\newblock
{\BBOQ}\APACrefatitle {{Equations of motion for a small rigid sphere in a
  nonuniform flow}} {{Equations of motion for a small rigid sphere in a
  nonuniform flow}}.{\BBCQ}
\newblock
\APACjournalVolNumPages{Physics of Fluids}{26}{}{883--889}.
\PrintBackRefs{\CurrentBib}

\bibitem [\protect \citeauthoryear {%
Meehl%
\ \protect \BOthers {.}}{%
Meehl%
\ \protect \BOthers {.}}{%
{\protect \APACyear {2007}}%
}]{%
Meehl2007}
\APACinsertmetastar {%
Meehl2007}%
\begin{APACrefauthors}%
Meehl, G\BPBI A.%
, Stocker, T\BPBI F.%
, Collins, W\BPBI D.%
, Friedlingstein, P.%
, Gaye, A\BPBI T.%
, Gregory, J\BPBI M.%
\BDBL {}Zhao, Z\BHBI C.%
\end{APACrefauthors}%
\unskip\
\newblock
\APACrefYearMonthDay{2007}{}{}.
\newblock
{\BBOQ}\APACrefatitle {Global Climate Projections} {Global climate
  projections}.{\BBCQ}
\newblock
\BIn{} S.~Solomon\ \BOthers {.}\ (\BEDS), \APACrefbtitle {Climate Change 2007:
  The Physical Science Basis. Contribution of Working Group I to the Fourth
  Assessment Report of the Intergovernmental Panel on Climate Change} {Climate
  change 2007: The physical science basis. contribution of working group i to
  the fourth assessment report of the intergovernmental panel on climate
  change}\ (\BPGS\ 747--846).
\newblock
\APACaddressPublisher{Cambridge, United Kingdom and New York, NY,
  USA}{Cambridge University Press}.
\PrintBackRefs{\CurrentBib}

\bibitem [\protect \citeauthoryear {%
Meredith%
\ \protect \BOthers {.}}{%
Meredith%
\ \protect \BOthers {.}}{%
{\protect \APACyear {2019}}%
}]{%
Meredith2019PolarRegions}
\APACinsertmetastar {%
Meredith2019PolarRegions}%
\begin{APACrefauthors}%
Meredith, M.%
, Meier, M\BPBI F\BPBI G\BPBI P.%
, Murphy, E\BPBI J.%
, Thomas, D\BPBI R.%
, Makarov, K\BPBI N\BPBI M.%
, Turner, J.%
\BCBL {}\ \BBA {} et al.%
\end{APACrefauthors}%
\unskip\
\newblock
\APACrefYearMonthDay{2019}{}{}.
\newblock
{\BBOQ}\APACrefatitle {Polar Regions} {Polar regions}.{\BBCQ}
\newblock
\BIn{} H\BHBI O.~Pörtner\ \BOthers {.}\ (\BEDS), \APACrefbtitle {IPCC Special
  Report on the Ocean and Cryosphere in a Changing Climate} {Ipcc special
  report on the ocean and cryosphere in a changing climate}\ (\BPGS\ 203--320).
\newblock
\APACaddressPublisher{}{Cambridge University Press}.
\newblock
\begin{APACrefDOI} \doi{10.1017/9781009157964.005} \end{APACrefDOI}
\PrintBackRefs{\CurrentBib}

\bibitem [\protect \citeauthoryear {%
Michaelides%
}{%
Michaelides%
}{%
{\protect \APACyear {1997}}%
}]{%
Michaelides1997}
\APACinsertmetastar {%
Michaelides1997}%
\begin{APACrefauthors}%
Michaelides, E\BPBI E.%
\end{APACrefauthors}%
\unskip\
\newblock
\APACrefYearMonthDay{1997}{}{}.
\newblock
{\BBOQ}\APACrefatitle {{Review—the transient equation of motion for
  particles, bubbles, and droplets}} {{Review—the transient equation of
  motion for particles, bubbles, and droplets}}.{\BBCQ}
\newblock
\APACjournalVolNumPages{Journal of Fluids Engineering, Transactions of the
  ASME}{119}{2}{233--247}.
\newblock
\begin{APACrefDOI} \doi{10.1115/1.2819127} \end{APACrefDOI}
\PrintBackRefs{\CurrentBib}

\bibitem [\protect \citeauthoryear {%
Mott%
\ \protect \BOthers {.}}{%
Mott%
\ \protect \BOthers {.}}{%
{\protect \APACyear {2008}}%
}]{%
Mott2008}
\APACinsertmetastar {%
Mott2008}%
\begin{APACrefauthors}%
Mott, R.%
, Faure, F.%
, Lehning, M.%
, L{\"{o}}we, H.%
, Hynek, B.%
, Michlmayer, G.%
\BDBL {}Sch{\"{o}}ner, W.%
\end{APACrefauthors}%
\unskip\
\newblock
\APACrefYearMonthDay{2008}{}{}.
\newblock
{\BBOQ}\APACrefatitle {{Simulation of seasonal snow-cover distribution for
  glacierized sites on Sonnblick, Austria, with the Alpine3D model}}
  {{Simulation of seasonal snow-cover distribution for glacierized sites on
  Sonnblick, Austria, with the Alpine3D model}}.{\BBCQ}
\newblock
\APACjournalVolNumPages{Annals of Glaciology}{49}{}{155--160}.
\newblock
\begin{APACrefDOI} \doi{10.3189/172756408787814924} \end{APACrefDOI}
\PrintBackRefs{\CurrentBib}

\bibitem [\protect \citeauthoryear {%
Mott%
, Schirmer%
, Bavay%
, Gr{\"{u}}newald%
\BCBL {}\ \BBA {} Lehning%
}{%
Mott%
\ \protect \BOthers {.}}{%
{\protect \APACyear {2010}}%
}]{%
Mott2010}
\APACinsertmetastar {%
Mott2010}%
\begin{APACrefauthors}%
Mott, R.%
, Schirmer, M.%
, Bavay, M.%
, Gr{\"{u}}newald, T.%
\BCBL {}\ \BBA {} Lehning, M.%
\end{APACrefauthors}%
\unskip\
\newblock
\APACrefYearMonthDay{2010}{dec}{}.
\newblock
{\BBOQ}\APACrefatitle {{Understanding snow-transport processes shaping the
  mountain snow-cover}} {{Understanding snow-transport processes shaping the
  mountain snow-cover}}.{\BBCQ}
\newblock
\APACjournalVolNumPages{The Cryosphere}{4}{4}{545--559}.
\newblock
\begin{APACrefDOI} \doi{10.5194/tc-4-545-2010} \end{APACrefDOI}
\PrintBackRefs{\CurrentBib}

\bibitem [\protect \citeauthoryear {%
Mott%
, Vionnet%
\BCBL {}\ \BBA {} Gr{\"{u}}newald%
}{%
Mott%
\ \protect \BOthers {.}}{%
{\protect \APACyear {2018}}%
}]{%
Mott2018}
\APACinsertmetastar {%
Mott2018}%
\begin{APACrefauthors}%
Mott, R.%
, Vionnet, V.%
\BCBL {}\ \BBA {} Gr{\"{u}}newald, T.%
\end{APACrefauthors}%
\unskip\
\newblock
\APACrefYearMonthDay{2018}{}{}.
\newblock
{\BBOQ}\APACrefatitle {{The Seasonal Snow Cover Dynamics: Review on Wind-Driven
  Coupling Processes}} {{The Seasonal Snow Cover Dynamics: Review on
  Wind-Driven Coupling Processes}}.{\BBCQ}
\newblock
\APACjournalVolNumPages{Frontiers in Earth Science}{6}{December}{}.
\newblock
\begin{APACrefDOI} \doi{10.3389/feart.2018.00197} \end{APACrefDOI}
\PrintBackRefs{\CurrentBib}

\bibitem [\protect \citeauthoryear {%
Nemes%
, Dasari%
, Hong%
, Guala%
\BCBL {}\ \BBA {} Coletti%
}{%
Nemes%
\ \protect \BOthers {.}}{%
{\protect \APACyear {2017}}%
}]{%
Nemes2017}
\APACinsertmetastar {%
Nemes2017}%
\begin{APACrefauthors}%
Nemes, A.%
, Dasari, T.%
, Hong, J.%
, Guala, M.%
\BCBL {}\ \BBA {} Coletti, F.%
\end{APACrefauthors}%
\unskip\
\newblock
\APACrefYearMonthDay{2017}{}{}.
\newblock
{\BBOQ}\APACrefatitle {{Snowflakes in the atmospheric surface layer:
  Observation of particle-turbulence dynamics}} {{Snowflakes in the atmospheric
  surface layer: Observation of particle-turbulence dynamics}}.{\BBCQ}
\newblock
\APACjournalVolNumPages{Journal of Fluid Mechanics}{814}{}{592--613}.
\newblock
\begin{APACrefDOI} \doi{10.1017/jfm.2017.13} \end{APACrefDOI}
\PrintBackRefs{\CurrentBib}

\bibitem [\protect \citeauthoryear {%
Nielsen%
}{%
Nielsen%
}{%
{\protect \APACyear {1993}}%
}]{%
Nielsen1993}
\APACinsertmetastar {%
Nielsen1993}%
\begin{APACrefauthors}%
Nielsen, P.%
\end{APACrefauthors}%
\unskip\
\newblock
\APACrefYearMonthDay{1993}{}{}.
\newblock
{\BBOQ}\APACrefatitle {{Turbulence Effects on the Settling of Suspended
  Particles}} {{Turbulence Effects on the Settling of Suspended
  Particles}}.{\BBCQ}
\newblock
\APACjournalVolNumPages{Journal of Sedimentary Research}{63}{5}{835--838}.
\newblock
\begin{APACrefDOI} \doi{10.1306/D4267C1C-2B26-11D7-8648000102C1865D}
  \end{APACrefDOI}
\PrintBackRefs{\CurrentBib}

\bibitem [\protect \citeauthoryear {%
Nishimura%
\ \protect \BOthers {.}}{%
Nishimura%
\ \protect \BOthers {.}}{%
{\protect \APACyear {2024}}%
}]{%
Nishimura2024}
\APACinsertmetastar {%
Nishimura2024}%
\begin{APACrefauthors}%
Nishimura, K.%
, Nemoto, M.%
, Ito, Y.%
, Omiya, S.%
, Shimoyama, K.%
\BCBL {}\ \BBA {} Niiya, H.%
\end{APACrefauthors}%
\unskip\
\newblock
\APACrefYearMonthDay{2024}{}{}.
\newblock
{\BBOQ}\APACrefatitle {{Elucidation of Spatiotemporal structures from
  high-resolution blowing snow observations}} {{Elucidation of Spatiotemporal
  structures from high-resolution blowing snow observations}}.{\BBCQ}
\newblock
\APACjournalVolNumPages{The Cryosphere}{18}{}{4775--4786}.
\PrintBackRefs{\CurrentBib}

\bibitem [\protect \citeauthoryear {%
Nishimura%
, Sugiura%
, Nemoto%
\BCBL {}\ \BBA {} Maeno%
}{%
Nishimura%
\ \protect \BOthers {.}}{%
{\protect \APACyear {1998}}%
}]{%
Nishimura1998}
\APACinsertmetastar {%
Nishimura1998}%
\begin{APACrefauthors}%
Nishimura, K.%
, Sugiura, K.%
, Nemoto, M.%
\BCBL {}\ \BBA {} Maeno, N.%
\end{APACrefauthors}%
\unskip\
\newblock
\APACrefYearMonthDay{1998}{}{}.
\newblock
{\BBOQ}\APACrefatitle {{Measurements and numerical simulations of snow-particle
  saltation}} {{Measurements and numerical simulations of snow-particle
  saltation}}.{\BBCQ}
\newblock
\APACjournalVolNumPages{Annals of Glaciology}{26}{}{184--190}.
\PrintBackRefs{\CurrentBib}

\bibitem [\protect \citeauthoryear {%
Oettinger%
, Ault%
, Stone%
\BCBL {}\ \BBA {} Haller%
}{%
Oettinger%
\ \protect \BOthers {.}}{%
{\protect \APACyear {2018}}%
}]{%
Oettinger2018}
\APACinsertmetastar {%
Oettinger2018}%
\begin{APACrefauthors}%
Oettinger, D.%
, Ault, J\BPBI T.%
, Stone, H\BPBI A.%
\BCBL {}\ \BBA {} Haller, G.%
\end{APACrefauthors}%
\unskip\
\newblock
\APACrefYearMonthDay{2018}{}{}.
\newblock
{\BBOQ}\APACrefatitle {{Invisible Anchors Trap Particles in Branching
  Junctions}} {{Invisible Anchors Trap Particles in Branching
  Junctions}}.{\BBCQ}
\newblock
\APACjournalVolNumPages{Physical Review Letters}{121}{5}{54502}.
\newblock
\begin{APACrefDOI} \doi{10.1103/PhysRevLett.121.054502} \end{APACrefDOI}
\PrintBackRefs{\CurrentBib}

\bibitem [\protect \citeauthoryear {%
Okaze%
, Niiya%
\BCBL {}\ \BBA {} Nishimura%
}{%
Okaze%
\ \protect \BOthers {.}}{%
{\protect \APACyear {2018}}%
}]{%
Okaze2018}
\APACinsertmetastar {%
Okaze2018}%
\begin{APACrefauthors}%
Okaze, T.%
, Niiya, H.%
\BCBL {}\ \BBA {} Nishimura, K.%
\end{APACrefauthors}%
\unskip\
\newblock
\APACrefYearMonthDay{2018}{}{}.
\newblock
{\BBOQ}\APACrefatitle {{Development of a large-eddy simulation coupled with
  Lagrangian snow transport model}} {{Development of a large-eddy simulation
  coupled with Lagrangian snow transport model}}.{\BBCQ}
\newblock
\APACjournalVolNumPages{Journal of Wind Engineering and Industrial
  Aerodynamics}{183}{September}{35--43}.
\newblock
\begin{APACrefDOI} \doi{10.1016/j.jweia.2018.09.027} \end{APACrefDOI}
\PrintBackRefs{\CurrentBib}

\bibitem [\protect \citeauthoryear {%
Perlman%
, Burns%
, Li%
\BCBL {}\ \BBA {} Meneveau%
}{%
Perlman%
\ \protect \BOthers {.}}{%
{\protect \APACyear {2007}}%
}]{%
Perlman2007}
\APACinsertmetastar {%
Perlman2007}%
\begin{APACrefauthors}%
Perlman, E.%
, Burns, R.%
, Li, Y.%
\BCBL {}\ \BBA {} Meneveau, C.%
\end{APACrefauthors}%
\unskip\
\newblock
\APACrefYearMonthDay{2007}{}{}.
\newblock
{\BBOQ}\APACrefatitle {Data Exploration of Turbulence Simulations using a
  Database Cluster} {Data exploration of turbulence simulations using a
  database cluster}.{\BBCQ}
\newblock
\BIn{} \APACrefbtitle {Proceedings of the Supercomputing SC07.} {Proceedings of
  the supercomputing sc07.}
\newblock
\APACaddressPublisher{}{ACM, IEEE}.
\newblock
\begin{APACrefDOI} \doi{10.1145/1362622.1362623} \end{APACrefDOI}
\PrintBackRefs{\CurrentBib}

\bibitem [\protect \citeauthoryear {%
Platt%
}{%
Platt%
}{%
{\protect \APACyear {1997}}%
}]{%
Platt1997}
\APACinsertmetastar {%
Platt1997}%
\begin{APACrefauthors}%
Platt, C\BPBI M\BPBI R.%
\end{APACrefauthors}%
\unskip\
\newblock
\APACrefYearMonthDay{1997}{}{}.
\newblock
{\BBOQ}\APACrefatitle {{A parameterization of the visible extinction
  coefficient of ice clouds in terms of the ice/water content}} {{A
  parameterization of the visible extinction coefficient of ice clouds in terms
  of the ice/water content}}.{\BBCQ}
\newblock
\APACjournalVolNumPages{Journal of the Atmospheric
  Sciences}{54}{16}{2083--2098}.
\newblock
\begin{APACrefDOI} \doi{10.1175/1520-0469(1997)054<2083:apotve>2.0.co;2}
  \end{APACrefDOI}
\PrintBackRefs{\CurrentBib}

\bibitem [\protect \citeauthoryear {%
Pomeroy%
\ \BBA {} Gray%
}{%
Pomeroy%
\ \BBA {} Gray%
}{%
{\protect \APACyear {1990}}%
}]{%
Pomeroy1990}
\APACinsertmetastar {%
Pomeroy1990}%
\begin{APACrefauthors}%
Pomeroy, J.%
\BCBT {}\ \BBA {} Gray, D.%
\end{APACrefauthors}%
\unskip\
\newblock
\APACrefYearMonthDay{1990}{}{}.
\newblock
{\BBOQ}\APACrefatitle {{Saltation of snow}} {{Saltation of snow}}.{\BBCQ}
\newblock
\APACjournalVolNumPages{Water resources research}{26}{7}{1583--1594}.
\PrintBackRefs{\CurrentBib}

\bibitem [\protect \citeauthoryear {%
Pomeroy%
\ \protect \BOthers {.}}{%
Pomeroy%
\ \protect \BOthers {.}}{%
{\protect \APACyear {2007}}%
}]{%
Pomeroy2007}
\APACinsertmetastar {%
Pomeroy2007}%
\begin{APACrefauthors}%
Pomeroy, J.%
, Gray, D.%
, Brown, T.%
, Hedstrom, N\BPBI R.%
, Quinton, W\BPBI L.%
, Granger, R\BPBI J.%
\BCBL {}\ \BBA {} Carey, S\BPBI K.%
\end{APACrefauthors}%
\unskip\
\newblock
\APACrefYearMonthDay{2007}{}{}.
\newblock
{\BBOQ}\APACrefatitle {{The cold regions hydrological model: a platform for
  basing process representation and model structure on physical evidence}}
  {{The cold regions hydrological model: a platform for basing process
  representation and model structure on physical evidence}}.{\BBCQ}
\newblock
\APACjournalVolNumPages{Hydrological Processes}{21}{}{2650--2667}.
\newblock
\begin{APACrefDOI} \doi{10.1002/hyp} \end{APACrefDOI}
\PrintBackRefs{\CurrentBib}

\bibitem [\protect \citeauthoryear {%
Pomeroy%
, Gray%
\BCBL {}\ \BBA {} Landine%
}{%
Pomeroy%
\ \protect \BOthers {.}}{%
{\protect \APACyear {1993}}%
}]{%
Pomeroy1993}
\APACinsertmetastar {%
Pomeroy1993}%
\begin{APACrefauthors}%
Pomeroy, J.%
, Gray, D.%
\BCBL {}\ \BBA {} Landine, P.%
\end{APACrefauthors}%
\unskip\
\newblock
\APACrefYearMonthDay{1993}{}{}.
\newblock
{\BBOQ}\APACrefatitle {{The prairie blowing snow model: characteristics,
  validation, operation}} {{The prairie blowing snow model: characteristics,
  validation, operation}}.{\BBCQ}
\newblock
\APACjournalVolNumPages{Journal of Hydrology}{144}{}{165--192}.
\PrintBackRefs{\CurrentBib}

\bibitem [\protect \citeauthoryear {%
Prokop%
}{%
Prokop%
}{%
{\protect \APACyear {2008}}%
}]{%
Prokop2008}
\APACinsertmetastar {%
Prokop2008}%
\begin{APACrefauthors}%
Prokop, A.%
\end{APACrefauthors}%
\unskip\
\newblock
\APACrefYearMonthDay{2008}{}{}.
\newblock
{\BBOQ}\APACrefatitle {{Assessing the applicability of terrestrial laser
  scanning for spatial snow depth measurements}} {{Assessing the applicability
  of terrestrial laser scanning for spatial snow depth measurements}}.{\BBCQ}
\newblock
\APACjournalVolNumPages{Cold Regions Science and Technology}{54}{3}{155--163}.
\newblock
\begin{APACrefDOI} \doi{10.1016/j.coldregions.2008.07.002} \end{APACrefDOI}
\PrintBackRefs{\CurrentBib}

\bibitem [\protect \citeauthoryear {%
Prokop%
}{%
Prokop%
}{%
{\protect \APACyear {2009}}%
}]{%
Prokop2009}
\APACinsertmetastar {%
Prokop2009}%
\begin{APACrefauthors}%
Prokop, A.%
\end{APACrefauthors}%
\unskip\
\newblock
\APACrefYearMonthDay{2009}{}{}.
\newblock
{\BBOQ}\APACrefatitle {{Terrestrial laser scanning for snow depth observations:
  An update on technical developments and applications}} {{Terrestrial laser
  scanning for snow depth observations: An update on technical developments and
  applications}}.{\BBCQ}
\newblock
\APACjournalVolNumPages{ISSW 09 - International Snow Science Workshop,
  Proceedings}{}{}{192--196}.
\PrintBackRefs{\CurrentBib}

\bibitem [\protect \citeauthoryear {%
Prokop%
\ \BBA {} Panholzer%
}{%
Prokop%
\ \BBA {} Panholzer%
}{%
{\protect \APACyear {2009}}%
}]{%
Prokop2009b}
\APACinsertmetastar {%
Prokop2009b}%
\begin{APACrefauthors}%
Prokop, A.%
\BCBT {}\ \BBA {} Panholzer, H.%
\end{APACrefauthors}%
\unskip\
\newblock
\APACrefYearMonthDay{2009}{}{}.
\newblock
{\BBOQ}\APACrefatitle {{Assessing the capability of terrestrial laser scanning
  for monitoring slow moving landslides}} {{Assessing the capability of
  terrestrial laser scanning for monitoring slow moving landslides}}.{\BBCQ}
\newblock
\APACjournalVolNumPages{Natural Hazards and Earth System
  Science}{9}{6}{1921--1928}.
\newblock
\begin{APACrefDOI} \doi{10.5194/nhess-9-1921-2009} \end{APACrefDOI}
\PrintBackRefs{\CurrentBib}

\bibitem [\protect \citeauthoryear {%
Prokop%
\ \BBA {} Procter%
}{%
Prokop%
\ \BBA {} Procter%
}{%
{\protect \APACyear {2016}}%
}]{%
Prokop2016}
\APACinsertmetastar {%
Prokop2016}%
\begin{APACrefauthors}%
Prokop, A.%
\BCBT {}\ \BBA {} Procter, E\BPBI S.%
\end{APACrefauthors}%
\unskip\
\newblock
\APACrefYearMonthDay{2016}{}{}.
\newblock
{\BBOQ}\APACrefatitle {{A new methodology for planning snow drift fences in
  alpine terrain}} {{A new methodology for planning snow drift fences in alpine
  terrain}}.{\BBCQ}
\newblock
\APACjournalVolNumPages{Cold Regions Science and Technology}{132}{}{33--43}.
\newblock
\begin{APACrefDOI} \doi{10.1016/j.coldregions.2016.09.010} \end{APACrefDOI}
\PrintBackRefs{\CurrentBib}

\bibitem [\protect \citeauthoryear {%
Rees%
, Singh%
, Pardyjak%
\BCBL {}\ \BBA {} Garrett%
}{%
Rees%
\ \protect \BOthers {.}}{%
{\protect \APACyear {2021}}%
}]{%
Rees2021}
\APACinsertmetastar {%
Rees2021}%
\begin{APACrefauthors}%
Rees, K\BPBI N.%
, Singh, D\BPBI K.%
, Pardyjak, E\BPBI R.%
\BCBL {}\ \BBA {} Garrett, T\BPBI J.%
\end{APACrefauthors}%
\unskip\
\newblock
\APACrefYearMonthDay{2021}{}{}.
\newblock
{\BBOQ}\APACrefatitle {{Mass and density of individual frozen hydrometeors}}
  {{Mass and density of individual frozen hydrometeors}}.{\BBCQ}
\newblock
\APACjournalVolNumPages{Atmospheric Chemistry and
  Physics}{21}{18}{14235--14250}.
\newblock
\begin{APACrefDOI} \doi{10.5194/acp-21-14235-2021} \end{APACrefDOI}
\PrintBackRefs{\CurrentBib}

\bibitem [\protect \citeauthoryear {%
Salesky%
, Giometto%
, Chamecki%
, Lehning%
\BCBL {}\ \BBA {} Parlange%
}{%
Salesky%
\ \protect \BOthers {.}}{%
{\protect \APACyear {2019}}%
}]{%
Salesky2019}
\APACinsertmetastar {%
Salesky2019}%
\begin{APACrefauthors}%
Salesky, S\BPBI T.%
, Giometto, M\BPBI G.%
, Chamecki, M.%
, Lehning, M.%
\BCBL {}\ \BBA {} Parlange, M\BPBI B.%
\end{APACrefauthors}%
\unskip\
\newblock
\APACrefYearMonthDay{2019}{}{}.
\newblock
{\BBOQ}\APACrefatitle {The transport and deposition of heavy particles in
  complex terrain: insights from an Eulerian model for large eddy simulation}
  {The transport and deposition of heavy particles in complex terrain: insights
  from an eulerian model for large eddy simulation}.{\BBCQ}
\newblock
\APACjournalVolNumPages{arXiv preprint arXiv:1903.03521}{}{}{}.
\newblock
\begin{APACrefURL} \url{https://doi.org/10.48550/arXiv.1903.03521}
  \end{APACrefURL}
\PrintBackRefs{\CurrentBib}

\bibitem [\protect \citeauthoryear {%
Saloranta%
}{%
Saloranta%
}{%
{\protect \APACyear {2012}}%
}]{%
Saloranta2012}
\APACinsertmetastar {%
Saloranta2012}%
\begin{APACrefauthors}%
Saloranta, T\BPBI M.%
\end{APACrefauthors}%
\unskip\
\newblock
\APACrefYearMonthDay{2012}{}{}.
\newblock
{\BBOQ}\APACrefatitle {{Simulating snow maps for Norway: Description and
  statistical evaluation of the seNorge snow model}} {{Simulating snow maps for
  Norway: Description and statistical evaluation of the seNorge snow
  model}}.{\BBCQ}
\newblock
\APACjournalVolNumPages{Cryosphere}{6}{6}{1323--1337}.
\newblock
\begin{APACrefDOI} \doi{10.5194/tc-6-1323-2012} \end{APACrefDOI}
\PrintBackRefs{\CurrentBib}

\bibitem [\protect \citeauthoryear {%
Schmidt%
}{%
Schmidt%
}{%
{\protect \APACyear {1980}}%
}]{%
Schmidt1980}
\APACinsertmetastar {%
Schmidt1980}%
\begin{APACrefauthors}%
Schmidt, R.%
\end{APACrefauthors}%
\unskip\
\newblock
\APACrefYearMonthDay{1980}{}{}.
\newblock
{\BBOQ}\APACrefatitle {{Threshold Wind-Speeds And Elastic Impact In Snow
  Transport}} {{Threshold Wind-Speeds And Elastic Impact In Snow
  Transport}}.{\BBCQ}
\newblock
\APACjournalVolNumPages{Journal of Glaciology}{26}{94}{453--467}.
\PrintBackRefs{\CurrentBib}

\bibitem [\protect \citeauthoryear {%
Schneiderbauer%
\ \BBA {} Prokop%
}{%
Schneiderbauer%
\ \BBA {} Prokop%
}{%
{\protect \APACyear {2011}}%
}]{%
Schneiderbauer2011}
\APACinsertmetastar {%
Schneiderbauer2011}%
\begin{APACrefauthors}%
Schneiderbauer, S.%
\BCBT {}\ \BBA {} Prokop, A.%
\end{APACrefauthors}%
\unskip\
\newblock
\APACrefYearMonthDay{2011}{jun}{}.
\newblock
{\BBOQ}\APACrefatitle {{The atmospheric snow-transport model: SnowDrift3D}}
  {{The atmospheric snow-transport model: SnowDrift3D}}.{\BBCQ}
\newblock
\APACjournalVolNumPages{Journal of Glaciology}{57}{203}{526--542}.
\newblock
\begin{APACrefDOI} \doi{10.3189/002214311796905677} \end{APACrefDOI}
\PrintBackRefs{\CurrentBib}

\bibitem [\protect \citeauthoryear {%
Sch{\"{o}}n%
, Naaim-Bouvet%
, Vionnet%
\BCBL {}\ \BBA {} Prokop%
}{%
Sch{\"{o}}n%
\ \protect \BOthers {.}}{%
{\protect \APACyear {2018}}%
}]{%
Schon2018}
\APACinsertmetastar {%
Schon2018}%
\begin{APACrefauthors}%
Sch{\"{o}}n, P.%
, Naaim-Bouvet, F.%
, Vionnet, V.%
\BCBL {}\ \BBA {} Prokop, A.%
\end{APACrefauthors}%
\unskip\
\newblock
\APACrefYearMonthDay{2018}{}{}.
\newblock
{\BBOQ}\APACrefatitle {{Merging a terrain-based parameter with blowing snow
  fluxes for assessing snow redistribution in alpine terrain}} {{Merging a
  terrain-based parameter with blowing snow fluxes for assessing snow
  redistribution in alpine terrain}}.{\BBCQ}
\newblock
\APACjournalVolNumPages{Cold Regions Science and Technology}{155}{November
  2017}{161--173}.
\newblock
\begin{APACrefDOI} \doi{10.1016/j.coldregions.2018.08.002} \end{APACrefDOI}
\PrintBackRefs{\CurrentBib}

\bibitem [\protect \citeauthoryear {%
Sch{\"{o}}n%
\ \protect \BOthers {.}}{%
Sch{\"{o}}n%
\ \protect \BOthers {.}}{%
{\protect \APACyear {2015}}%
}]{%
Schon2015}
\APACinsertmetastar {%
Schon2015}%
\begin{APACrefauthors}%
Sch{\"{o}}n, P.%
, Prokop, A.%
, Vionnet, V.%
, Guyomarc'h, G.%
, Naaim-Bouvet, F.%
\BCBL {}\ \BBA {} Heiser, M.%
\end{APACrefauthors}%
\unskip\
\newblock
\APACrefYearMonthDay{2015}{}{}.
\newblock
{\BBOQ}\APACrefatitle {{Improving a terrain-based parameter for the assessment
  of snow depths with TLS data in the Col du Lac Blanc area}} {{Improving a
  terrain-based parameter for the assessment of snow depths with TLS data in
  the Col du Lac Blanc area}}.{\BBCQ}
\newblock
\APACjournalVolNumPages{Cold Regions Science and Technology}{114}{}{15--26}.
\newblock
\begin{APACrefDOI} \doi{10.1016/j.coldregions.2015.02.005} \end{APACrefDOI}
\PrintBackRefs{\CurrentBib}

\bibitem [\protect \citeauthoryear {%
Schweizer%
, Jamieson%
\BCBL {}\ \BBA {} Schneebeli%
}{%
Schweizer%
\ \protect \BOthers {.}}{%
{\protect \APACyear {2003}}%
}]{%
Schweizer2003}
\APACinsertmetastar {%
Schweizer2003}%
\begin{APACrefauthors}%
Schweizer, J.%
, Jamieson, J\BPBI B.%
\BCBL {}\ \BBA {} Schneebeli, M.%
\end{APACrefauthors}%
\unskip\
\newblock
\APACrefYearMonthDay{2003}{}{}.
\newblock
{\BBOQ}\APACrefatitle {{Snow avalanche formation}} {{Snow avalanche
  formation}}.{\BBCQ}
\newblock
\APACjournalVolNumPages{Reviews of Geophysics}{41}{4}{1--25}.
\newblock
\begin{APACrefDOI} \doi{10.1029/2002RG000123} \end{APACrefDOI}
\PrintBackRefs{\CurrentBib}

\bibitem [\protect \citeauthoryear {%
Singh%
, Donovan%
, Pardyjak%
\BCBL {}\ \BBA {} Garrett%
}{%
Singh%
\ \protect \BOthers {.}}{%
{\protect \APACyear {2021}}%
}]{%
Singh2021}
\APACinsertmetastar {%
Singh2021}%
\begin{APACrefauthors}%
Singh, D\BPBI K.%
, Donovan, S.%
, Pardyjak, E\BPBI R.%
\BCBL {}\ \BBA {} Garrett, T\BPBI J.%
\end{APACrefauthors}%
\unskip\
\newblock
\APACrefYearMonthDay{2021}{}{}.
\newblock
{\BBOQ}\APACrefatitle {{A differential emissivity imaging technique for
  measuring hydrometeor mass and type}} {{A differential emissivity imaging
  technique for measuring hydrometeor mass and type}}.{\BBCQ}
\newblock
\APACjournalVolNumPages{Atmospheric Measurement
  Techniques}{14}{11}{6973--6990}.
\newblock
\begin{APACrefDOI} \doi{10.5194/amt-14-6973-2021} \end{APACrefDOI}
\PrintBackRefs{\CurrentBib}

\bibitem [\protect \citeauthoryear {%
Singh%
, Pardyjak%
\BCBL {}\ \BBA {} Garrett%
}{%
Singh%
\ \protect \BOthers {.}}{%
{\protect \APACyear {2023}}%
}]{%
Singh2023}
\APACinsertmetastar {%
Singh2023}%
\begin{APACrefauthors}%
Singh, D\BPBI K.%
, Pardyjak, E\BPBI R.%
\BCBL {}\ \BBA {} Garrett, T\BPBI J.%
\end{APACrefauthors}%
\unskip\
\newblock
\APACrefYearMonthDay{2023}{}{}.
\newblock
{\BBOQ}\APACrefatitle {{A universal scaling law for Lagrangian snowflake
  accelerations in atmospheric turbulence}} {{A universal scaling law for
  Lagrangian snowflake accelerations in atmospheric turbulence}}.{\BBCQ}
\newblock
\APACjournalVolNumPages{Physics of Fluids}{35}{12}{}.
\newblock
\begin{APACrefDOI} \doi{10.1063/5.0173359} \end{APACrefDOI}
\PrintBackRefs{\CurrentBib}

\bibitem [\protect \citeauthoryear {%
Singh%
, Pardyjak%
\BCBL {}\ \BBA {} Garrett%
}{%
Singh%
\ \protect \BOthers {.}}{%
{\protect \APACyear {2024}}%
}]{%
Singh2024}
\APACinsertmetastar {%
Singh2024}%
\begin{APACrefauthors}%
Singh, D\BPBI K.%
, Pardyjak, E\BPBI R.%
\BCBL {}\ \BBA {} Garrett, T\BPBI J.%
\end{APACrefauthors}%
\unskip\
\newblock
\APACrefYearMonthDay{2024}{}{}.
\newblock
{\BBOQ}\APACrefatitle {{Time-resolved measurements of the densities of
  individual frozen hydrometeors and fresh snowfall}} {{Time-resolved
  measurements of the densities of individual frozen hydrometeors and fresh
  snowfall}}.{\BBCQ}
\newblock
\APACjournalVolNumPages{Atmospheric Measurement
  Techniques}{17}{15}{4581--4598}.
\newblock
\begin{APACrefDOI} \doi{10.5194/amt-17-4581-2024} \end{APACrefDOI}
\PrintBackRefs{\CurrentBib}

\bibitem [\protect \citeauthoryear {%
Tom%
\ \BBA {} Bragg%
}{%
Tom%
\ \BBA {} Bragg%
}{%
{\protect \APACyear {2019}}%
}]{%
Tom2019}
\APACinsertmetastar {%
Tom2019}%
\begin{APACrefauthors}%
Tom, J.%
\BCBT {}\ \BBA {} Bragg, A\BPBI D.%
\end{APACrefauthors}%
\unskip\
\newblock
\APACrefYearMonthDay{2019}{}{}.
\newblock
{\BBOQ}\APACrefatitle {{Multiscale preferential sweeping of particles settling
  in turbulence}} {{Multiscale preferential sweeping of particles settling in
  turbulence}}.{\BBCQ}
\newblock
\APACjournalVolNumPages{Journal of Fluid Mechanics}{871}{}{244--270}.
\newblock
\begin{APACrefDOI} \doi{10.1017/jfm.2019.337} \end{APACrefDOI}
\PrintBackRefs{\CurrentBib}

\bibitem [\protect \citeauthoryear {%
Vaughan%
\ \protect \BOthers {.}}{%
Vaughan%
\ \protect \BOthers {.}}{%
{\protect \APACyear {2013}}%
}]{%
IPCC2013Chapter4}
\APACinsertmetastar {%
IPCC2013Chapter4}%
\begin{APACrefauthors}%
Vaughan, D\BPBI G.%
, Comiso, J\BPBI C.%
, Allison, I.%
, Carrasco, J.%
, Kaser, G.%
, Kwok, R.%
\BDBL {}Zhang, T.%
\end{APACrefauthors}%
\unskip\
\newblock
\APACrefYearMonthDay{2013}{}{}.
\newblock
{\BBOQ}\APACrefatitle {Chapter 4: Observations: Cryosphere} {Chapter 4:
  Observations: Cryosphere}.{\BBCQ}
\newblock
\BIn{} T\BPBI F.~Stocker\ \BOthers {.}\ (\BEDS), \APACrefbtitle {Climate Change
  2013: The Physical Science Basis. Contribution of Working Group I to the
  Fifth Assessment Report of the Intergovernmental Panel on Climate Change}
  {Climate change 2013: The physical science basis. contribution of working
  group i to the fifth assessment report of the intergovernmental panel on
  climate change}\ (\BPGS\ 317--382).
\newblock
\APACaddressPublisher{Cambridge, United Kingdom and New York, NY,
  USA}{Cambridge University Press}.
\PrintBackRefs{\CurrentBib}

\bibitem [\protect \citeauthoryear {%
Vionnet%
\ \protect \BOthers {.}}{%
Vionnet%
\ \protect \BOthers {.}}{%
{\protect \APACyear {2021}}%
}]{%
Vionnet2021}
\APACinsertmetastar {%
Vionnet2021}%
\begin{APACrefauthors}%
Vionnet, V.%
, Marsh, C\BPBI B.%
, Menounos, B.%
, Gascoin, S.%
, Wayand, N\BPBI E.%
, Shea, J.%
\BDBL {}Pomeroy, J\BPBI W.%
\end{APACrefauthors}%
\unskip\
\newblock
\APACrefYearMonthDay{2021}{}{}.
\newblock
{\BBOQ}\APACrefatitle {{Multi-scale snowdrift-permitting modelling of mountain
  snowpack}} {{Multi-scale snowdrift-permitting modelling of mountain
  snowpack}}.{\BBCQ}
\newblock
\APACjournalVolNumPages{Cryosphere}{15}{2}{743--769}.
\newblock
\begin{APACrefDOI} \doi{10.5194/tc-15-743-2021} \end{APACrefDOI}
\PrintBackRefs{\CurrentBib}

\bibitem [\protect \citeauthoryear {%
Vionnet%
\ \protect \BOthers {.}}{%
Vionnet%
\ \protect \BOthers {.}}{%
{\protect \APACyear {2014}}%
}]{%
Vionnet2014}
\APACinsertmetastar {%
Vionnet2014}%
\begin{APACrefauthors}%
Vionnet, V.%
, Martin, E.%
, Masson, V.%
, Guyomarc'h, G.%
, Naaim-Bouvet, F.%
, Prokop, A.%
\BDBL {}Lac, C.%
\end{APACrefauthors}%
\unskip\
\newblock
\APACrefYearMonthDay{2014}{mar}{}.
\newblock
{\BBOQ}\APACrefatitle {{Simulation of wind-induced snow transport and
  sublimation in alpine terrain using a fully coupled snowpack/atmosphere
  model}} {{Simulation of wind-induced snow transport and sublimation in alpine
  terrain using a fully coupled snowpack/atmosphere model}}.{\BBCQ}
\newblock
\APACjournalVolNumPages{The Cryosphere}{8}{2}{395--415}.
\newblock
\begin{APACrefDOI} \doi{10.5194/tc-8-395-2014} \end{APACrefDOI}
\PrintBackRefs{\CurrentBib}

\bibitem [\protect \citeauthoryear {%
Voth%
\ \BBA {} Soldati%
}{%
Voth%
\ \BBA {} Soldati%
}{%
{\protect \APACyear {2017}}%
}]{%
Voth2017}
\APACinsertmetastar {%
Voth2017}%
\begin{APACrefauthors}%
Voth, G\BPBI A.%
\BCBT {}\ \BBA {} Soldati, A.%
\end{APACrefauthors}%
\unskip\
\newblock
\APACrefYearMonthDay{2017}{}{}.
\newblock
{\BBOQ}\APACrefatitle {{Anisotropic Particles in Turbulence}} {{Anisotropic
  Particles in Turbulence}}.{\BBCQ}
\newblock
\APACjournalVolNumPages{Annual Review of Fluid Mechanics}{49}{}{249--276}.
\newblock
\begin{APACrefDOI} \doi{10.1146/annurev-fluid-010816-060135} \end{APACrefDOI}
\PrintBackRefs{\CurrentBib}

\bibitem [\protect \citeauthoryear {%
Wang%
\ \BBA {} Maxey%
}{%
Wang%
\ \BBA {} Maxey%
}{%
{\protect \APACyear {1993}}%
}]{%
Wang1993}
\APACinsertmetastar {%
Wang1993}%
\begin{APACrefauthors}%
Wang, L\BPBI P.%
\BCBT {}\ \BBA {} Maxey, M\BPBI R.%
\end{APACrefauthors}%
\unskip\
\newblock
\APACrefYearMonthDay{1993}{}{}.
\newblock
{\BBOQ}\APACrefatitle {{Settling velocity and concentration distribution of
  heavy particles in homogeneous isotropic turbulence}} {{Settling velocity and
  concentration distribution of heavy particles in homogeneous isotropic
  turbulence}}.{\BBCQ}
\newblock
\APACjournalVolNumPages{Journal of Fluid Mechanics}{256}{3}{27--68}.
\newblock
\begin{APACrefDOI} \doi{10.1017/S0022112093002708} \end{APACrefDOI}
\PrintBackRefs{\CurrentBib}

\bibitem [\protect \citeauthoryear {%
Winstral%
, Elder%
\BCBL {}\ \BBA {} Davis%
}{%
Winstral%
\ \protect \BOthers {.}}{%
{\protect \APACyear {2002}}%
}]{%
Winstral2002a}
\APACinsertmetastar {%
Winstral2002a}%
\begin{APACrefauthors}%
Winstral, A.%
, Elder, K.%
\BCBL {}\ \BBA {} Davis, R\BPBI E.%
\end{APACrefauthors}%
\unskip\
\newblock
\APACrefYearMonthDay{2002}{}{}.
\newblock
{\BBOQ}\APACrefatitle {{Spatial snow modeling of wind-redistributed snow using
  terrain-based parameters}} {{Spatial snow modeling of wind-redistributed snow
  using terrain-based parameters}}.{\BBCQ}
\newblock
\APACjournalVolNumPages{Journal of Hydrometeorology}{3}{5}{524--538}.
\newblock
\begin{APACrefDOI} \doi{10.1175/1525-7541(2002)003<0524:SSMOWR>2.0.CO;2}
  \end{APACrefDOI}
\PrintBackRefs{\CurrentBib}

\bibitem [\protect \citeauthoryear {%
Winstral%
, Jonas%
\BCBL {}\ \BBA {} Helbig%
}{%
Winstral%
\ \protect \BOthers {.}}{%
{\protect \APACyear {2017}}%
}]{%
Winstral2017}
\APACinsertmetastar {%
Winstral2017}%
\begin{APACrefauthors}%
Winstral, A.%
, Jonas, T.%
\BCBL {}\ \BBA {} Helbig, N.%
\end{APACrefauthors}%
\unskip\
\newblock
\APACrefYearMonthDay{2017}{}{}.
\newblock
{\BBOQ}\APACrefatitle {{Statistical downscaling of gridded wind speed data
  using local topography}} {{Statistical downscaling of gridded wind speed data
  using local topography}}.{\BBCQ}
\newblock
\APACjournalVolNumPages{Journal of Hydrometeorology}{18}{2}{335--348}.
\newblock
\begin{APACrefDOI} \doi{10.1175/JHM-D-16-0054.1} \end{APACrefDOI}
\PrintBackRefs{\CurrentBib}

\bibitem [\protect \citeauthoryear {%
Winstral%
\ \BBA {} Marks%
}{%
Winstral%
\ \BBA {} Marks%
}{%
{\protect \APACyear {2002}}%
}]{%
Winstral2002b}
\APACinsertmetastar {%
Winstral2002b}%
\begin{APACrefauthors}%
Winstral, A.%
\BCBT {}\ \BBA {} Marks, D.%
\end{APACrefauthors}%
\unskip\
\newblock
\APACrefYearMonthDay{2002}{}{}.
\newblock
{\BBOQ}\APACrefatitle {{Simulating wind fields and snow redistribution using
  terrain-based parameters to model snow accumulation and melt over a semi-arid
  mountain catchment}} {{Simulating wind fields and snow redistribution using
  terrain-based parameters to model snow accumulation and melt over a semi-arid
  mountain catchment}}.{\BBCQ}
\newblock
\APACjournalVolNumPages{Hydrological Processes}{16}{18}{3585--3603}.
\newblock
\begin{APACrefDOI} \doi{10.1002/hyp.1238} \end{APACrefDOI}
\PrintBackRefs{\CurrentBib}

\bibitem [\protect \citeauthoryear {%
Wood%
}{%
Wood%
}{%
{\protect \APACyear {1995}}%
}]{%
Wood1995}
\APACinsertmetastar {%
Wood1995}%
\begin{APACrefauthors}%
Wood, N.%
\end{APACrefauthors}%
\unskip\
\newblock
\APACrefYearMonthDay{1995}{}{}.
\newblock
{\BBOQ}\APACrefatitle {{The onset of separation in neutral, turbulent flow over
  hills}} {{The onset of separation in neutral, turbulent flow over
  hills}}.{\BBCQ}
\newblock
\APACjournalVolNumPages{Boundary-Layer Meteorology}{76}{1-2}{137--164}.
\newblock
\begin{APACrefDOI} \doi{10.1007/BF00710894} \end{APACrefDOI}
\PrintBackRefs{\CurrentBib}

\bibitem [\protect \citeauthoryear {%
Zaki%
}{%
Zaki%
}{%
{\protect \APACyear {2013}}%
}]{%
Zaki2013}
\APACinsertmetastar {%
Zaki2013}%
\begin{APACrefauthors}%
Zaki, T\BPBI A.%
\end{APACrefauthors}%
\unskip\
\newblock
\APACrefYearMonthDay{2013}{}{}.
\newblock
{\BBOQ}\APACrefatitle {From streaks to spots and on to turbulence: exploring
  the dynamics of boundary layer transition} {From streaks to spots and on to
  turbulence: exploring the dynamics of boundary layer transition}.{\BBCQ}
\newblock
\APACjournalVolNumPages{Flow, Turbulence and Combustion}{91}{3}{451--473}.
\newblock
\begin{APACrefDOI} \doi{10.1007/s10494-013-9483-6} \end{APACrefDOI}
\PrintBackRefs{\CurrentBib}

\end{thebibliography}
%


%
%
%
%
%

\end{document}